\documentclass[11pt]{article}
\usepackage[a4paper,top=2.4cm,bottom=2.6cm,left=2.2cm,right=2.2cm]{geometry}
\usepackage[utf8]{inputenc}
\usepackage[expansion=false]{microtype}
\usepackage{graphicx}
\usepackage{amsmath}
\usepackage{xcolor}
\usepackage[most]{tcolorbox}
\usepackage{titlesec}
\usepackage{caption}
\usepackage{enumitem}
\usepackage{fancyhdr}
\usepackage{url}
\usepackage{marvosym}
\usepackage[hidelinks]{hyperref}
\definecolor{npjblue}{RGB}{0,0,0}
\definecolor{boxframe}{RGB}{0,0,0}
\definecolor{boxbg}{RGB}{245,245,245}
\definecolor{rulegrey}{RGB}{120,120,120}

\titleformat{\section}{\bfseries\large}{}{0pt}{}
\titlespacing*{\section}{0pt}{16pt plus 2pt}{7pt}
\titleformat{\subsection}{\bfseries\normalsize}{}{0pt}{}
\titlespacing*{\subsection}{0pt}{12pt plus 2pt}{5pt}

\DeclareCaptionLabelFormat{pipe}{#1 #2 $|$}
\newcommand{\sfcite}[1]{\textsuperscript{#1}}

\newlist{refs}{enumerate}{1}
\setlist[refs]{label=\arabic*.,leftmargin=1.8em,itemsep=1.2pt,parsep=0pt,font=\footnotesize}

\begin{document}
\thispagestyle{plain}

\begin{flushleft}
{\footnotesize\sffamily\bfseries\color{npjblue}PERSPECTIVE}\\[2pt]
{\color{rulegrey!60}\rule{\textwidth}{0.5pt}}\\[10pt]
{\bfseries\LARGE Towards a physics-informed multiscale digital twin for precision medicine in Alzheimer's disease\par}
\vspace{10pt}
{\normalsize Aida Nonn\textsuperscript{1,\Letter}, Elma Kerz\textsuperscript{2}, Daniel Wiechmann\textsuperscript{2,3} \& Paul Edison\textsuperscript{4}\par}
\vspace{7pt}
{\footnotesize
\textsuperscript{1}Ostbayerische Technische Hochschule Regensburg, Regensburg, Germany.
\textsuperscript{2}Exaia Technologies GmbH, Aachen, Germany.
\textsuperscript{3}University of Amsterdam, Amsterdam, The Netherlands.
\textsuperscript{4}Imperial College London, London, UK.\\[2pt]
\textsuperscript{\Letter}Correspondence and requests for materials should be addressed to Aida Nonn (\href{mailto:aida.nonn@oth-regensburg.de}{aida.nonn@oth-regensburg.de}).\par}
\end{flushleft}

\vspace{6pt}
\begin{tcolorbox}[colback=boxbg,colframe=boxbg,arc=1.5mm,left=3mm,right=3mm,top=2.5mm,bottom=2.5mm]
{\sffamily\bfseries\small\color{npjblue} Abstract}\\[3pt]
\small Deciphering the drivers of brain ageing and neurodegeneration, and explaining heterogeneity in individual trajectories, remains a central challenge for precision medicine. We propose PIM-BrainTwin, a physics-informed multiscale digital twin for Alzheimer's disease that draws on materials-science principles---fatigue-like depletion, safety margins and critical transitions---to quantify residual compensatory capacity and system stability. Designed as an open, modular and federated platform, it enables alternative mechanistic hypotheses to be formalised, compared and refined.
\end{tcolorbox}
\vspace{2pt}

\section*{Digital twins for precision medicine in brain health}

Deciphering the fundamental drivers of brain ageing and neurodegeneration, and explaining substantial heterogeneity in individual trajectories, remains one of the defining challenges of twenty-first-century neuroscience. Precision medicine has advanced this agenda by linking genomic, molecular, clinical, environmental and lifestyle information to patient stratification, risk prediction and treatment selection.\sfcite{1--3} High-resolution molecular profiling and neuroimaging, electrophysiology and behavioural phenotyping, combined with recent advances in artificial intelligence and digital biomarker technologies, now provide longitudinal, multiscale and ecologically sensitive readouts of individual brain trajectories.\sfcite{4--7} Yet the key methodological bottleneck remains the integration of these fragmentary, heterogeneous multiscale and multi-temporal observations---spanning molecular, circuit and systems levels across spatial and temporal scales of many orders of magnitude, from nanometre-scale protein aggregation unfolding over years, through millisecond network dynamics, to day-to-day fluctuations in real-world behaviour---into patient-specific, dynamically updatable models of disease mechanisms, individual trajectories and intervention responses. Digital twins provide this computational layer for precision brain medicine by continuously synchronising multimodal patient data with mechanistic, generative or hybrid simulations to estimate latent disease states, forecast trajectory evolution and evaluate therapeutic strategies in silico.\sfcite{8--10}

In this Perspective, we propose a physics-informed multiscale digital-twin framework for the brain --- henceforth PIM-BrainTwin --- that integrates concepts from computational mechanics, biomechanics and materials science into precision brain-health modelling. Its central premise is that brain health is actively maintained by interacting vascular, metabolic, immune, glial, pathological, network and behavioural systems that absorb and redistribute biological load, thereby preserving function even as pathology accumulates. In health, these systems interact dynamically to preserve homeostasis and support recovery from perturbation. When homeostatic processes are impaired, protective and compensatory mechanisms redistribute functional demand across cellular, vascular, metabolic and network systems to maintain physiological stability. This compensatory capacity declines gradually during normal ageing; in neurodegeneration, disease-specific processes accelerate and self-reinforce the same loss of reserve. In both cases the erosion often remains clinically silent until reserves are substantially depleted and the coupled system approaches decompensation. Materials science and multiscale computational mechanics have spent decades developing rigorous mathematical tools precisely for this class of problem: how complex dynamical systems progressively lose function under repeated loading, how damage accumulates across scales, how residual function is preserved through compensatory redistribution of load, and how local failures can eventually propagate into system-wide instability.\sfcite{11--17}

These methods therefore offer a principled foundation for modelling brain ageing and neurodegeneration. When embedded in computational simulations, they allow the interacting processes of transport, clearance, haemodynamics, tissue mechanics and biological stress--recovery to be represented as coupled physical and biological constraints on individual trajectories. Concepts such as fatigue-like depletion, hysteresis, resilience, safety margins, critical transition and multiscale computational homogenisation thereby become quantitative tools for estimating remaining compensatory capacity, identifying dominant routes to decompensation, and constraining digital-twin predictions to biologically and physically plausible regimes. Readers are referred to Box 1 for definitions of these concepts and their application within the proposed framework.

\begin{tcolorbox}[breakable,enhanced,colback=boxbg,colframe=boxframe,boxrule=0.6pt,arc=1.5mm,left=3mm,right=3mm,top=2.5mm,bottom=2.5mm,title={\sffamily\bfseries Box 1 $|$ Operational definitions of key context-specific concepts},fonttitle=\normalsize,coltitle=white,colbacktitle=boxframe,before skip=10pt,after skip=10pt]
\small\setlength{\parskip}{4.5pt}\setlength{\parindent}{0pt}

\textbf{Physics-informed, coupled multiscale digital-twin.} A dynamically updated computational representation of an individual's brain that integrates longitudinal multimodal observations with mechanistic models constrained by physical principles, experimentally derived parameters and empirically supported causal mechanisms. \emph{Physics-informed} means that model behaviour is constrained by governing equations, conservation principles and biological knowledge, while individual susceptibility and biological uncertainty are represented through individual-specific parameters, heterogeneous initial conditions, stochastic processes and observation models that are continuously updated from each individual's data. \emph{Coupled} means that interacting biological and physical subsystems influence one another's evolution through explicitly represented causal mechanisms, feedback, compensation and amplification. \emph{Multiscale} means that processes operating across cellular, tissue, whole-brain network and behavioural scales are linked through explicit scale-bridging relationships. \emph{Digital twin} means that the computational representation is repeatedly synchronised with new observations, allowing its latent states, parameters and forecasts to evolve with the individual.

\textbf{Multiscale computational homogenisation.} A mathematical framework for deriving effective tissue-scale properties from explicitly represented heterogeneous microstructure. In brain modelling, it relates cerebral microvascular and tissue microstructure to continuum parameters governing perfusion, interstitial transport, poroelastic behaviour and tissue mechanics.

\textbf{Complex dynamical system.} A system composed of multiple interacting components whose states evolve over time through coupled, often nonlinear relationships. In the brain, causal interactions among cellular, vascular, immune, metabolic, network and behavioural processes generate feedback, compensation, adaptation and amplification, such that system-level trajectories emerge from their collective dynamics rather than from any single component. The behaviour of the system depends on its current state, prior history and external perturbations, and may exhibit distinct dynamical regimes and transitions between them.

\textbf{Biological load.} The cumulative demand imposed on a biological system by internal and external perturbations. In brain health, it reflects the magnitude, duration, recurrence and interaction of pathological, inflammatory, vascular, metabolic, mechanical and environmental stressors that engage compensatory and recovery processes over time.

\textbf{Latent state and observation model.} A latent state is an underlying biological or physiological condition that evolves over time but is not measured directly. An observation model specifies how imaging, molecular, electrophysiological, behavioural and other measurements relate to one or more latent states, thereby separating the dynamics of the biological system from their modality-specific manifestations.

\textbf{Compensatory capacity.} The ability of interacting biological systems to maintain physiological stability and functional integrity despite accumulating perturbation or loss of underlying reserve. In the brain, it reflects the combined capacity of cellular, vascular, metabolic, immune and network processes to buffer stress, redistribute biological demand and support recovery over time.

\textbf{Fatigue-like depletion.} A progressive, history-dependent reduction in compensatory capacity under repeated or sustained biological load, formalised by continuum-damage-type evolution equations in which residual vulnerability accumulates when recovery between successive exposures is incomplete. At the cellular level this is expressed through internal state variables (e.g. residual proteostatic, metabolic or oxidative load) whose rates of accumulation and recovery are set by experimentally constrained parameters such as clearance capacity, energetic efficiency and damage thresholds.

\textbf{Hysteresis.} History dependence in which recovery follows a trajectory distinct from decline, such that the current state reflects prior biological load as well as present conditions.

\textbf{Safety margin.} The difference between current compensatory capacity and the minimum capacity required to maintain a stable, compensated state under prevailing biological load. A larger margin indicates greater tolerance to additional perturbation, whereas a narrowing margin indicates increasing vulnerability to loss of compensation.

\textbf{Critical transition}. A qualitative shift in dynamical regime that occurs when compensatory capacity falls below the level required to maintain stability, causing the loss of a previously stable operating state. Beyond this threshold, recovery is incomplete or absent, and the coupled multiscale processes move into a persistently degraded regime. In the present architecture, decompensation corresponds to the biological transition.

\end{tcolorbox}
PIM-BrainTwin is introduced as an open-source, modular framework for the research community, within which alternative mechanistic accounts can be formalised, compared and iteratively refined. Alzheimer's disease serves as the primary use case, providing a stringent setting characterised by a prolonged preclinical phase, marked inter-individual heterogeneity, well-characterised molecular and imaging biomarkers, the emergence of disease-modifying interventions, and imperfect correspondence between clinical diagnosis and neuropathology.\sfcite{18--22} Within this setting we begin by testing progressive loss of glial adaptive capacity as the initial biological hypothesis, representing it as a latent state that couples vascular, inflammatory--clearance, metabolic, pathological and network-level processes.

\section*{Alzheimer's disease as a multiscale complex dynamical system}

Alzheimer's disease represents a spatially distributed, multiscale pathological process in which genetic, molecular, cellular, vascular, immune, metabolic and systems-level mechanisms interact through nonlinear feedback loops.\sfcite{23--30} Contemporary disease-primer and mechanistic-review accounts reinforce this view by framing amyloid dysregulation, tau pathology, synaptic and neuronal injury, oxidative--metabolic stress, neuroinflammation, vascular dysfunction, impaired clearance and network degeneration as interacting processes rather than isolated explanatory pathways.\sfcite{29,30} Its multiscale and heterarchical organisation requires complex dynamical-systems and systems-biology frameworks capable of capturing emergent behaviours, nonlinear dynamics and state transitions across biological scales.\sfcite{24,25,31--33} Integrative disease-progression models have begun to formalise this view by combining subtype inference, connectome-constrained propagation, patient-specific biomarker dynamics and multimodal trajectory modelling.\sfcite{34--39} The remaining challenge is to extend these models beyond retrospective reconstruction towards prospective estimation of latent state, compensatory stability, mechanism-specific vulnerability and future trajectory.

To address this challenge, the first testable biological hypothesis examined within the PIM-BrainTwin framework is the progressive loss of glial adaptive capacity. This choice is motivated by the integrative position of glial systems in Alzheimer's disease. Astrocytes support neurovascular-unit and blood--brain-barrier function, metabolic buffering, synaptic and ionic homeostasis, tissue mechanics and aquaporin-4-dependent perivascular clearance, while microglia regulate inflammatory--clearance dynamics through innate immune sensing, disease-associated phagocytic programmes, cytokine-mediated signalling and stage-dependent interactions with amyloid and tau pathology.\sfcite{40--54} Glial adaptive capacity is thus represented as a biologically grounded latent state embedded within a common state-space formulation of vascular, inflammatory--clearance, neuronal--metabolic, pathological and functional-network dynamics. This glial-centred instantiation provides a mechanistically coherent starting point while retaining the capacity to quantify how vascular regulation, inflammatory--clearance dynamics, neuronal--metabolic support, pathology propagation and functional-network physiology differentially contribute to compensatory failure across individuals, datasets and model variants.\sfcite{29,30,48--50,55}

Building on multiscale dynamical-systems, critical-transition and tipping-point formulations of Alzheimer's disease and neurodegeneration,\sfcite{31--33} we develop one concrete realisation of PIM-BrainTwin in which progression from preclinical or prodromal pathology to clinical Alzheimer's disease is modelled as the erosion of compensatory stability and eventual loss of a stable, compensated state within a coupled neurovascular--glial--metabolic system. In this realisation, the latent system state is represented as

\begin{equation*} X_t=\left(A_t,\,V_t,\,I_t,\,M_t,\,P_t,\,F_t\right)^{\!\top}, \end{equation*}

where $A_t$ denotes glial adaptive capacity, $V_t$ vascular--haemodynamic support, $I_t$ inflammatory--clearance state, $M_t$ neuronal--metabolic support, $P_t$ amyloid--tau burden and $F_t$ functional-network physiology. System stability is made explicit through the safety margin

\begin{equation*} S_t = A_t - A_t^{\mathrm{crit}}. \end{equation*}

Here, $A_t^{\mathrm{crit}}$ denotes the minimum adaptive capacity required to maintain a stable, compensated state under the prevailing biological load. Accordingly, $S_t$ quantifies the signed difference between current adaptive capacity and this critical threshold. Physics-informed simulation links cellular stress--recovery processes, tissue-scale transport and mechanics, connectome-constrained pathology propagation and multimodal observations. This enables latent physiological states to be estimated, tracked and forecast under mechanistic constraints and through multiscale parameter transfer. In this formulation, the primary object of inference shifts from pathology burden alone to the remaining compensatory capacity of the coupled system, the mechanisms that deplete it and the proximity of an individual trajectory to the transition from prodromal pathology to clinical Alzheimer's disease. Readers are referred to Supplementary Note 1 and Supplementary Table 1 for a detailed mapping of biological domains onto state variables, measurements and observation equations.

Genetic susceptibility is incorporated as time-invariant hierarchical priors on compensatory capacity, repair--depletion kinetics and subsystem-specific interaction strengths; readers are referred to Supplementary Note 2 and Supplementary Table 2 for details. Consistent with cellular-phase accounts of Alzheimer's disease, APOE genotype, genome-wide polygenic risk and cell-type-weighted risk scores are treated not merely as predictors of disease probability, but as modifiers of the dynamical parameters that govern glial, vascular, inflammatory--clearance and neuronal--metabolic trajectories.\sfcite{24,28,56--59} In contrast, cell-state transcriptomics provides a time-varying molecular observation: gene-expression programmes change with cell type, ageing, environmental exposure and disease stage, and single-nucleus atlases show progressive shifts from homeostatic toward stress-associated glial and neuronal states across the Alzheimer's continuum.\sfcite{25--27,60,61} These transcriptomic signatures inform prior distributions and observation maps for latent physiological states---including the glial adaptive-capacity state $A_t$---rather than serving as static omics classifiers. Vascular, inflammatory--clearance, neuronal--metabolic, pathological and network measurements further constrain their corresponding latent states, while speech and cognition supply higher-frequency observations of functional expression. Cognitive reserve is represented within the observation model that links functional-network physiology to measured cognition and behaviour, so that preserved cognitive performance despite declining network physiology becomes informative about residual compensatory capacity rather than an unexplained deviation from the underlying disease trajectory.\sfcite{51,62--65}

\section*{A physics-informed, coupled multiscale framework}

Figure 1 provides a schematic representation of the physics-informed, coupled multiscale organisation of PIM-BrainTwin. The framework links cellular stress--recovery dynamics, tissue-scale transport and mechanics, connectome-constrained propagation and individual-level observations through explicit cross-scale parameter transfer.\sfcite{8--17,34--39}

At the micro scale, cellular and neurovascular-unit systems are exposed to inflammatory, metabolic, amyloid--tau--clearance and mechanical--haemodynamic stressors to estimate repair kinetics, depletion kinetics, vulnerability parameters and persistent loss of recoverable capacity. These measurements parameterise glial adaptive capacity, $A_t$, as a stress-history-dependent state variable shaped by fatigue-like depletion and incomplete recovery.\sfcite{40--54}

At the meso scale, statistical-volume-element (SVE) homogenisation translates microstructural and biomechanical properties---including endfoot geometry, aquaporin-4 polarity, perivascular architecture, basement-membrane properties and vascular compliance---into effective tissue coefficients governing transport, clearance, haemodynamic loading and mechanical strain. This parameter-passing step is central: physical quantities such as effective transport, clearance, compliance, porosity, stiffness, haemodynamic loading and mechanical stress constrain the space of biologically admissible individual-level trajectories.\sfcite{13,16,17,52,66} By restricting inference to trajectories that satisfy experimentally calibrated biological and physical constraints, PIM-BrainTwin reduces the effective dimensionality of the estimation problem and allows sparse individual observations to be combined with cross-scale parameters and population-level priors.

At the macro scale, these coefficients constrain an individual-specific digital twin defined on the structural connectome, where $A_t$, $V_t$, $I_t$, $M_t$, $P_t$ and $F_t$ evolve as coupled nonlinear dynamics with individual-specific parameters. Connectome-constrained propagation models provide the basis for representing spatial amyloid--tau dynamics within this whole-brain system.\sfcite{14,15,35--39} Personalisation occurs through Bayesian state estimation: MRI, PET, fluid biomarkers, transcriptomic profiles, high-density electroencephalography, speech and cognition constrain posterior estimates of latent states and parameters rather than serving as direct readouts of the state itself.\sfcite{8--10} Vascular and metabolic observations are especially important because early cerebral-blood-flow reduction, blood--brain-barrier dysfunction, oxidative stress and glucose-metabolic impairment can alter both available adaptive capacity and the capacity required to maintain compensation.\sfcite{48--50}

Artificial intelligence, particularly machine-learning and deep-learning methods, provides the data-modelling layer that makes PIM-BrainTwin operational across heterogeneous clinical modalities. In imaging, these methods support representation learning, segmentation, harmonisation and extraction of multimodal MRI, PET and connectome-derived features.\sfcite{7,67--71} In electrophysiology and digital phenotyping, they derive structured measures of network dynamics, cognition, speech and behaviour from high-dimensional or high-frequency signals.\sfcite{6,72--77} In clinical prediction and trial-facing stratification, machine-learning models integrate multimodal clinical, cognitive, imaging, voice and electronic-health-record features to support diagnostic, prognostic and enrichment tasks in dementia.\sfcite{78--81} In omics and cell-state data, they help map transcriptomic and genetic patterns onto biologically interpretable priors and observation functions.\sfcite{25--28,56,82} Finally, physics-informed machine learning, operator learning and symbolic regression can identify coupling terms or accelerate simulation while preserving the mechanistic constraints of the state-space model.\sfcite{39,83--85} Together, these methods convert raw multimodal observations into structured inputs for estimating latent state, safety margin, trajectory evolution and intervention response.

In this framework, personalised predictive modelling means estimating the stability of an individual coupled system. Each new observation updates the latent trajectory, refines the safety margin $S_t$, identifies which depletion pathways most strongly drive movement towards clinical transition and enables intervention scenarios to be evaluated in silico.\sfcite{31--33} This mechanistic personalisation estimates how an individual's coupled glial, vascular, inflammatory, metabolic, pathological and network dynamics are expected to evolve, with subtype assignment and risk stratification treated as secondary summaries of the inferred trajectory.\sfcite{34--39}

\begin{figure*}[p]
\centering
\includegraphics[width=\textwidth]{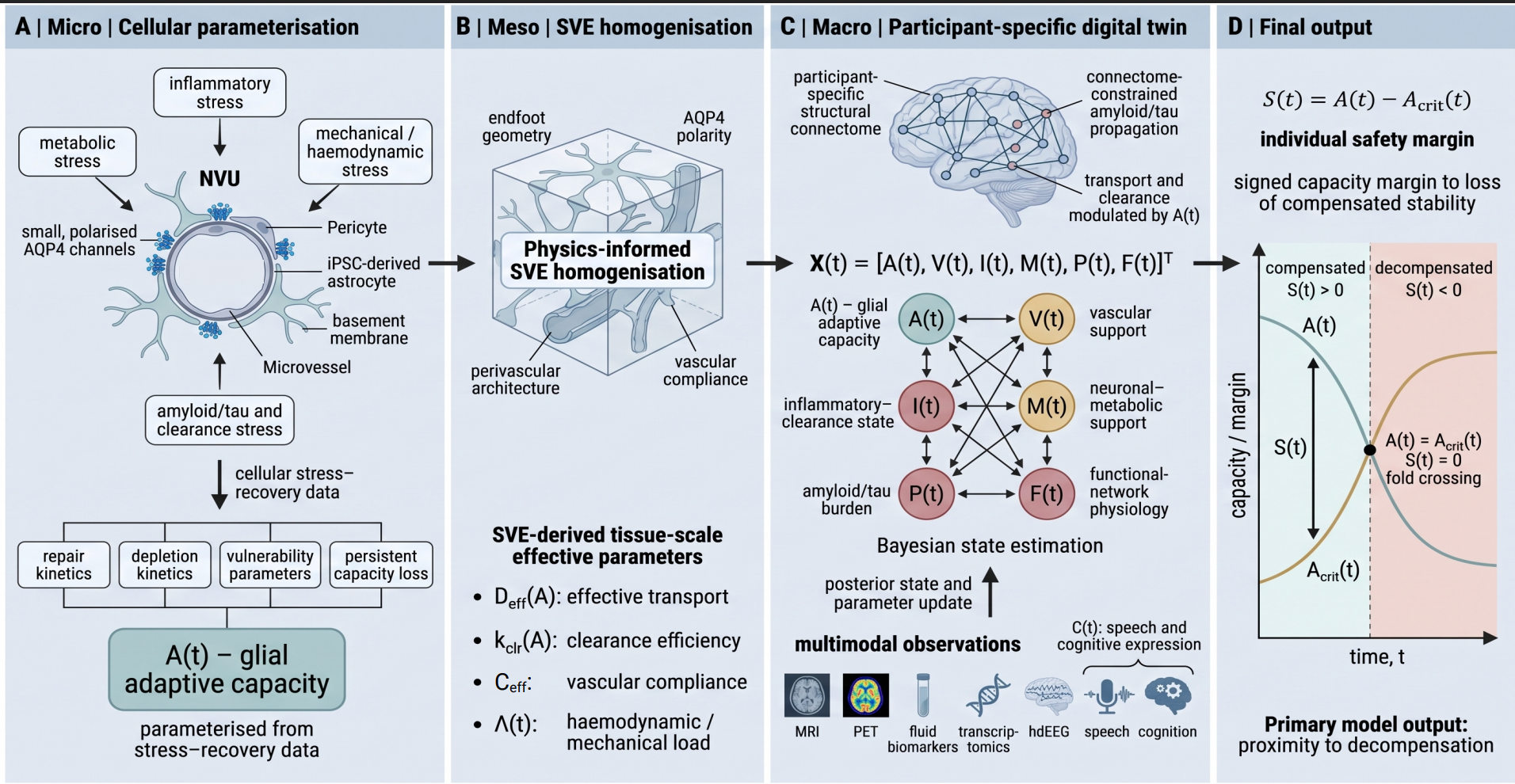}
\caption{\textbf{Physics-informed, coupled multiscale organisation of PIM-BrainTwin.} The figure shows how cellular perturbation data, tissue-scale effective parameters, connectome-constrained pathology propagation and longitudinal multimodal observations are integrated within an individual-specific state-space model. \textbf{A, Micro-scale calibration.} iPSC-derived astrocytes and neurovascular-unit systems are exposed to inflammatory, metabolic, amyloid--tau--clearance and mechanical--haemodynamic stressors to estimate repair kinetics, depletion kinetics, genotype-dependent vulnerability and persistent loss of recoverable capacity. These experiments provide experimentally grounded priors on glial adaptive capacity $A(t)$, inflammatory--clearance state $I(t)$, vascular--haemodynamic support $V(t)$ and metabolic support $M(t)$.\sfcite{40--54} \textbf{B, Meso-scale homogenisation.} Statistical-volume-element homogenisation of the neurovascular unit translates microstructural and biomechanical properties into tissue-level effective coefficients. Relevant quantities include astrocytic endfoot geometry, aquaporin-4 polarity, perivascular architecture, basement-membrane properties, vascular compliance and haemodynamic loading. These features determine effective parameters for perivascular or glymphatic clearance, tissue transport, mechanical compliance and stress transmission, thereby passing experimentally constrained micro-scale information to the individual-level model.\sfcite{13,16,17,52,66} \textbf{C, Macro-scale dynamical system.} The individual-specific digital twin is represented by the latent state vector $X(t)=[A(t),V(t),I(t),M(t),P(t),F(t)]^{\top}$, where $A(t)$ denotes glial adaptive capacity, $V(t)$ vascular--haemodynamic support, $I(t)$ inflammatory--clearance state, $M(t)$ neuronal--metabolic support, $P(t)$ amyloid--tau burden and $F(t)$ functional-network physiology. These variables evolve as a coupled nonlinear state-space system. Spatial amyloid--tau dynamics are represented by a connectome-constrained reaction--diffusion field on the structural connectome. The macro-scale model therefore combines local repair--depletion dynamics with network-mediated pathology propagation and whole-brain physiological coupling.\sfcite{14,15,35--39} \textbf{Multimodal observation and Bayesian updating.} Observations enter the model at complementary temporal resolutions. Low-frequency measurements include structural and functional imaging, amyloid and tau PET, FDG-PET, ASL-MRI and clinical phenotyping. Intermediate-resolution observations include transcriptomic, fluid-biomarker and inflammatory--clearance readouts. High-frequency observations include speech-derived digital biomarkers and cognitive or behavioural measures. Bayesian state estimation assimilates these data streams to update posterior estimates of latent states, parameters and future trajectories.\sfcite{6--10,25--28,56} \textbf{D, Precision brain-health output.} The primary derived quantity is the individual safety margin $S(t)=A(t)-A^{\mathrm{crit}}(t)$, defined as the distance between current adaptive capacity and the state-dependent boundary at which compensated dynamics lose stability. The model estimates how much compensatory capacity remains, which biological depletion pathways dominate the trajectory and how interventions may alter future evolution.\sfcite{31--33}}
\label{fig:1}
\end{figure*}

\section*{From disease-progression models to causal physics-informed digital twins}

Systems-biology studies, mechanism inventories and multi-omics atlases have substantially expanded the biological map of Alzheimer's disease, revealing convergent genetic, cellular, vascular, immune, metabolic, pathological and network domains.\sfcite{24--31,55,56} These resources provide essential mechanistic context, yet they primarily yield comprehensive characterisations of molecular pathways, cellular states and their interactions. The next requirement is predictive integration: models capable of converting multiscale observations into individual-level estimates of latent state, compensatory capacity, trajectory formation and intervention response.

Contemporary disease-progression modelling has moved the field beyond static diagnostic classification towards inference of trajectories. Phenomenological approaches---including event-based models, latent-time regression and subtype-and-stage inference---estimate temporal ordering, disease stage, subtype structure and biomarker trajectories.\sfcite{34,38} Pathophysiological approaches, such as network diffusion, epidemic spreading, reaction--diffusion and physics-informed equation-discovery methods, incorporate assumptions about regional vulnerability, network-mediated spread and hub susceptibility.\sfcite{14,35--39,83} Together, these approaches provide the temporal, spatial and phenotypic scaffold on which a mechanistic digital twin can build.

A critical methodological gap remains. Existing progression models rarely transmit experimentally calibrated cellular parameters through tissue-scale mechanical and transport coefficients into individual-specific whole-brain state equations. They also seldom encode physical constraints---transport, clearance, compliance and haemodynamic loading---as explicit bounds on biologically admissible trajectories.

PIM-BrainTwin is proposed as one concrete realisation that addresses this gap. It retains the temporal and spatial scaffolds of existing models while embedding them within an individual-specific multiscale state-space system. In this system, cellular stress--recovery kinetics, neurovascular-unit homogenisation, tissue-scale transport and mechanical coefficients, connectome-constrained propagation and multimodal observations jointly constrain estimates of compensatory capacity, depletion pathways and safety margin. Disease stage, subtype, regional pathology burden and biomarker trajectories are thereby interpreted as downstream phenotypic expressions of an underlying mechanistic trajectory rather than as isolated endpoints.

Formally, individual disease evolution is expressed as

\begin{equation*} \frac{\mathrm{d}X_t}{\mathrm{d}t} = f\!\left(X_t,\,u_t,\,p(x,t);\,\theta\right), \end{equation*}

where $X_t$ is the latent state vector, $u_t$ denotes perturbations or interventions, $p(x,t)$ represents spatially distributed amyloid--tau pathology and $\theta$ comprises individual- or population-level parameters. The function f encodes the coupled depletion, repair, propagation and feedback processes through which biological subsystems influence one another. Imaging, fluid biomarkers, electrophysiology, digital speech biomarkers update posterior estimates of latent states and parameters rather than serving as isolated endpoints. Candidate early-warning indicators---slowing recovery, rising autocorrelation, altered network stability or progressive narrowing of the safety margin---can then be evaluated as signatures of approaching loss of compensated dynamics.

The PIM-BrainTwin framework responds to recent calls for dynamical, multiscale and mechanistic modelling of Alzheimer's disease. Rollo and colleagues argue that the heterarchical organisation of Alzheimer's disease requires nonlinear models that integrate genetic, cellular, tissue-level and organ-level processes across biological and physical scales.\sfcite{31} Simons and colleagues frame clinical transition as a tipping-point process in which adaptive glial, immune and vascular functions preserve preclinical stability until self-reinforcing pathological cascades emerge.\sfcite{33} De Domenico and colleagues similarly argue that precision-medicine digital twins require mechanistic simulations grounded in explicit biological hypotheses and multiscale mechanisms.\sfcite{9} PIM-BrainTwin implements these principles in an individual-specific state-space framework that couples multiscale biological hypotheses with physical constraints, Bayesian state estimation, sensitivity analysis and formal model-credibility assessment.

The causal structure is specified through the state equations, coupling terms and parameterisation of the framework. Vascular-dominant, inflammatory--clearance, neuronal--metabolic, pathology-propagation and network-dysfunction trajectories correspond to distinct configurations of the same coupled system. Competing mechanistic hypotheses can therefore be compared by their ability to reproduce observed trajectories, predict future states and account for responses to targeted perturbations. Any causal interpretation remains conditional on the assumed framework structure and must be evaluated through sensitivity analysis, external validation and intervention data.

Physics-informed constraints operate across scales. Cellular stress--recovery experiments provide estimates of depletion and repair kinetics; meso-scale homogenisation maps microstructural and biomechanical properties onto effective tissue-level parameters for transport, clearance and compliance; connectome-constrained formulations describe the spatial propagation of amyloid and tau pathology; and Bayesian state estimation integrates multimodal observations into posterior estimates of latent states and future trajectories.\sfcite{9,13--17,23,39,66,83--85} This cross-scale parameter transfer gives biological and physical meaning to the latent states and restricts inferred trajectories to those consistent with experimental evidence and governing constraints.

The safety margin is one derived quantity within this broader framework. It estimates proximity to loss of compensated dynamics, while the wider aim is to identify the mechanisms that shape trajectory evolution, explain inter-individual heterogeneity, quantify nonlinear transition risk and simulate how targeted interventions may alter future outcomes. PIM-BrainTwin thus extends existing disease-progression approaches beyond staging, subtyping and propagation inference towards mechanistic state estimation, trajectory forecasting and mechanism-informed evaluation of intervention effects.

\subsection*{From digital replicas to precision brain-health scenarios}

PIM-BrainTwin is not intended to reproduce an individual brain in exhaustive detail. The relevant biological system spans molecular, cellular, tissue, network, behavioural and environmental scales, many of which remain only partially observable. Its value lies instead in disciplined abstraction: identifying the state variables, interactions and observations required to represent the dynamics relevant to a defined scientific or translational question. The framework therefore follows the central principle of precision-medicine digital twins---transparent, hypothesis-driven computational representations for evaluating biologically grounded scenarios rather than high-dimensional replicas of individuals.\sfcite{9,86,87}

This perspective changes the role of multimodal data. Imaging, fluid biomarkers, omics, electrophysiology, speech, cognition and real-world behavioural measures are not treated as parallel predictors. Each observation is linked to one or more latent states through an explicit observation function and contributes to updating the corresponding state or parameter estimates. Multimodal measurements thereby constrain a shared dynamical system in which biological hypotheses are represented through state equations, coupling terms and parameterisations, and inferred trajectories are evaluated for biological plausibility and credibility. This provides a common formal structure for integrating evidence from glial biology, vascular physiology, inflammation, metabolism, pathology propagation, network neuroscience and digital phenotyping.\sfcite{9,31,55}

The same framework supports several precision brain-health applications. Earlier detection becomes the estimation of latent trajectory change before conventional clinical endpoints become abnormal. Stratification becomes mechanism-informed, grouping individuals according to the processes that most strongly shape their trajectories rather than by syndrome or biomarker positivity alone. Changes in speech, cognition, PET, MRI, fluid biomarkers or electroencephalography can be interpreted as complementary manifestations of a shared underlying trajectory. Candidate therapies, neuromodulation strategies and combination interventions can be represented as perturbations to specific components of the coupled system, allowing their predicted effects on latent states, compensatory capacity and future trajectories to be compared.\sfcite{9,23,67,88}

\subsection*{Challenges and open questions}

Three challenges will determine whether PIM-BrainTwin can become scientifically credible and translationally useful. The first is identifiability. A coupled multiscale formulation may contain more parameters than can be estimated from any realistic dataset at the individual level. The framework must therefore distinguish parameters that can be inferred from individual observations from those that require cohort-level priors, experimental calibration or fixed structural assumptions. Physics-informed constraints, cellular stress--recovery experiments, cross-scale homogenisation and hierarchical Bayesian inference are essential because longitudinal observations will remain sparse, irregular and incomplete.\sfcite{66,84,85,89--91} These constraints do not replace informative longitudinal measurements, but they reduce dependence on dense individual data by limiting the admissible parameter and trajectory space.

The second challenge is measurement. Information-rich modalities, including amyloid and tau PET, TSPO-PET, advanced MRI, transcriptomics, fluid biomarkers and high-density electroencephalography, are costly and unevenly available across centres. More scalable measures, including speech, cognition, wearable sensors and real-world behavioural data, can increase sampling frequency but require rigorous analytical and clinical validation. Multicentre application therefore depends on harmonised acquisition protocols, preprocessing pipelines, quality-control procedures, observation functions, metadata standards and explicit documentation of missingness and measurement uncertainty.\sfcite{6,92--94}

The third challenge is validation and credibility. A digital twin cannot be evaluated by predictive accuracy alone. Evidence must be matched to a defined context of use, because exploratory hypothesis testing, retrospective trajectory reconstruction, trial enrichment, endpoint interpretation, treatment-response simulation and clinical decision support impose different requirements. Standardised verification, validation and uncertainty-quantification terminology is needed to distinguish numerical verification, empirical validation, sensitivity analysis, uncertainty propagation and overall credibility assessment.\sfcite{95} Risk-informed frameworks further relate the required evidence to the intended decision and the consequences of an incorrect prediction.\sfcite{96,97} PIM-BrainTwin must therefore be evaluated for robustness to missing data, sensitivity to structural assumptions, posterior calibration, biological plausibility of inferred states, external reproducibility and prospective validity of predicted intervention responses.

\subsection*{Roadmap and future directions}

A staged roadmap proceeds from open computational infrastructure to prospective translation. The first stage is to release the framework specification, state-variable definitions, governing equations, parameter priors, observation functions, synthetic datasets, validation protocols, sensitivity analyses and credibility-assessment criteria according to FAIR and reproducible-science principles.\sfcite{93--95} This aligns with recent calls for Alzheimer's disease modelling frameworks that make assumptions explicit, support multiscale hypothesis testing, enable shared data infrastructure and allow competing system-level accounts to be iteratively refined by the research community.\sfcite{31} PIM-BrainTwin is thus conceived as an open, modular and auditable framework in which alternative biological hypotheses---glial-centred, vascular, inflammatory, metabolic, pathological, network or reserve-centred---can be formalised, compared and revised. The mapping of major Alzheimer's disease hypotheses onto PIM-BrainTwin state variables, coupling terms and observation functions is summarised in Supplementary Note 3 and Supplementary Table 3. The computational architecture for capturing and integrating externally contributed data, assuring data and model validity, constructing and updating predictive models, and scaling the framework across centres and modalities---together with its technical feasibility under privacy and governance constraints---is detailed in Supplementary Note 4.

PIM-BrainTwin also provides a common modelling language for communities working at different biological scales. Genetic and functional-genomic studies can inform priors on baseline vulnerability, repair capacity and depletion-pathway parameters; single-cell atlases and molecular perturbation studies can refine observation functions and coupling terms; iPSC-derived astrocyte and neurovascular-unit experiments can estimate stress--recovery kinetics and persistent loss of recoverable capacity; vascular, inflammatory and metabolic imaging can constrain individual physiological states; speech-derived digital biomarkers, cognitive measures and electrophysiology can provide higher-frequency observations of functional expression; and disease-progression approaches can supply temporal, spatial and phenotypic scaffolds. Expressing these contributions within a shared coupled state-space formulation allows distributed mechanistic evidence to inform comparable estimates of latent state, safety margin, trajectory evolution and intervention effects.

The second stage is retrospective calibration and external evaluation in multimodal cohorts such as ADNI, OASIS, EPAD\sfcite{98--101} and BioFINDER. The third is prospective updating in deeply phenotyped longitudinal cohorts, in which repeated observations refine posterior estimates of latent states, parameters and future trajectories. The fourth is trial-facing evaluation for participant enrichment, endpoint interpretation and treatment-response analysis. Clinical decision support should remain downstream and contingent on identifiability, multicentre reproducibility, posterior calibration and prospective validation.\sfcite{89--91,95--97}

Although developed here for Alzheimer's disease, PIM-BrainTwin is extensible across diseases and mechanistic hypotheses. In Parkinson's disease, relevant latent processes might include dopaminergic--striatal dysfunction, $\alpha$-synuclein propagation, neuroinflammation, mitochondrial stress and network compensation.\sfcite{102--104} In other neurodegenerative diseases, vascular dysregulation, immune activation, protein propagation, synaptic failure or network instability may carry greater explanatory weight. The central requirement is that each hypothesis be expressed through a coupled state-space formulation with explicit state variables, observations, parameters, credibility criteria and intervention mappings.

\section*{Conclusion}

PIM-BrainTwin provides a framework for advancing precision brain-health modelling from multimodal association towards mechanistically constrained estimation of latent states, compensatory stability and future trajectories. By linking experimentally calibrated processes across cellular, tissue and whole-brain scales, it establishes a common formal structure in which competing biological hypotheses can be specified, updated with longitudinal observations and evaluated against explicit credibility criteria. Progressive loss of glial adaptive capacity provides one testable instantiation of this broader framework.

The immediate objective is to develop an open and auditable platform for determining which multiscale mechanisms are identifiable, which observations are most informative and how predicted intervention effects depend on the state of the coupled system. Progress will require rigorous cross-scale calibration, uncertainty quantification and external and prospective validation. Meeting these requirements would establish a principled route from fragmented multimodal measurements to dynamically updated, mechanism-informed estimates of individual brain-health trajectories.

\section*{References}
\begin{refs}
\item Collins, F. S. \& Varmus, H. A new initiative on precision medicine. \emph{N. Engl. J. Med.} \textbf{372}, 793--795 (2015).
\item Ashley, E. A. Towards precision medicine. \emph{Nat. Rev. Genet.} \textbf{17}, 507--522 (2016).
\item Hampel, H. et al. The foundation and architecture of precision medicine in neurology and psychiatry. \emph{Trends Neurosci.} \textbf{46}, 176--198 (2023).
\item Small, S. L. Precision neurology. \emph{Ageing Res. Rev.} \textbf{104}, 102632 (2025).
\item Topol, E. J. High-performance medicine: the convergence of human and artificial intelligence. \emph{Nat. Med.} \textbf{25}, 44--56 (2019).
\item Coravos, A., Khozin, S. \& Mandl, K. D. Developing and adopting safe and effective digital biomarkers to improve patient outcomes. \emph{npj Digit. Med.} \textbf{2}, 14 (2019).
\item Acosta, J. N., Falcone, G. J., Rajpurkar, P. \& Topol, E. J. Multimodal biomedical AI. \emph{Nat. Med.} \textbf{28}, 1773--1784 (2022).
\item Laubenbacher, R., Mehrad, B., Shmulevich, I. \& Trayanova, N. Digital twins in medicine. \emph{Nat. Comput. Sci.} \textbf{4}, 184--191 (2024).
\item De Domenico, M. et al. Challenges and opportunities for digital twins in precision medicine from a complex systems perspective. \emph{npj Digit. Med.} \textbf{8}, 37 (2025).
\item Wang, H. E. et al. Virtual brain twins: from basic neuroscience to clinical use. \emph{Natl Sci. Rev.} \textbf{11}, nwae079 (2024).
\item Suresh, S. \emph{Fatigue of Materials} 2nd edn (Cambridge Univ. Press, 1998).
\item Lemaitre, J. \& Chaboche, J.-L. \emph{Mechanics of Solid Materials} (Cambridge Univ. Press, 1990).
\item Geers, M. G. D., Kouznetsova, V. G. \& Brekelmans, W. A. M. Multi-scale computational homogenization: trends and challenges. \emph{J. Comput. Appl. Math.} \textbf{234}, 2175--2182 (2010).
\item Weickenmeier, J., Kuhl, E. \& Goriely, A. Multiphysics of prionlike diseases: progression and atrophy. \emph{Phys. Rev. Lett.} \textbf{121}, 158101 (2018).
\item Fornari, S., Schäfer, A., Jucker, M., Goriely, A. \& Kuhl, E. Prion-like spreading of Alzheimer's disease within the brain's connectome. \emph{J. R. Soc. Interface} \textbf{16}, 20190356 (2019).
\item Schäfer, A., Weickenmeier, J. \& Kuhl, E. The interplay of biochemical and biomechanical degeneration in Alzheimer's disease. \emph{Comput. Methods Appl. Mech. Eng.} \textbf{352}, 369--388 (2019).
\item Alber, M. et al. Integrating machine learning and multiscale modeling---perspectives, challenges, and opportunities in the biological, biomedical, and behavioral sciences. \emph{npj Digit. Med.} \textbf{2}, 115 (2019).
\item Jack, C. R. Jr et al. NIA-AA Research Framework: toward a biological definition of Alzheimer's disease. \emph{Alzheimers Dement.} \textbf{14}, 535--562 (2018).
\item Karran, E. \& Hardy, J. A critique of the drug discovery and phase 3 clinical programs targeting the amyloid hypothesis for Alzheimer disease. \emph{Ann. Neurol.} \textbf{76}, 185--205 (2014).
\item Cummings, J. et al. Alzheimer's disease drug development pipeline: 2022. \emph{Alzheimers Dement. Transl. Res. Clin. Interv.} \textbf{8}, e12295 (2022).
\item Zhang, Y., Chen, H., Li, R., Sterling, K. \& Song, W. Amyloid $\beta$-based therapy for Alzheimer's disease: challenges, successes and future. \emph{Signal Transduct. Target. Ther.} \textbf{8}, 248 (2023).
\item Beach, T. G., Monsell, S. E., Phillips, L. E. \& Kukull, W. Accuracy of the clinical diagnosis of Alzheimer disease at National Institute on Aging Alzheimer Disease Centers, 2005--2010. \emph{J. Neuropathol. Exp. Neurol.} \textbf{71}, 266--273 (2012).
\item Briggs, J. K. et al. Towards a physics informed digital twin to predict cerebral blood flow and cerebral vascular regulation. \emph{npj Digit. Med.} \textbf{9}, 428 (2026).
\item De Strooper, B. \& Karran, E. The cellular phase of Alzheimer's disease. \emph{Cell} \textbf{164}, 603--615 (2016).
\item Green, G. S. et al. Cellular communities reveal trajectories of brain ageing and Alzheimer's disease. \emph{Nature} \textbf{633}, 634--645 (2024).
\item Gabitto, M. I. et al. Integrated multimodal cell atlas of Alzheimer's disease. \emph{Nat. Neurosci.} \textbf{27}, 2366--2383 (2024).
\item Miyoshi, E. et al. Spatial and single-nucleus transcriptomic analysis of genetic and sporadic forms of Alzheimer's disease. \emph{Nat. Genet.} \textbf{56}, 2704--2717 (2024).
\item Bellenguez, C. et al. New insights into the genetic etiology of Alzheimer's disease and related dementias. \emph{Nat. Genet.} \textbf{54}, 412--436 (2022).
\item Knopman, D. S. et al. Alzheimer disease. \emph{Nat. Rev. Dis. Primers} \textbf{7}, 33 (2021).
\item Zhang, J. et al. Recent advances in Alzheimer's disease: mechanisms, clinical trials and new drug development strategies. \emph{Signal Transduct. Target. Ther.} \textbf{9}, 211 (2024).
\item Rollo, J. L., Crawford, J. \& Hardy, J. A dynamical systems approach for multiscale synthesis of Alzheimer's pathogenesis. \emph{Neuron} \textbf{111}, 2126--2139 (2023).
\item Scheffer, M. et al. Early-warning signals for critical transitions. \emph{Nature} \textbf{461}, 53--59 (2009).
\item Simons, M., Levin, J. \& Dichgans, M. Tipping points in neurodegeneration. \emph{Neuron} \textbf{111}, 2954--2968 (2023).
\item Young, A. L. et al. Uncovering the heterogeneity and temporal complexity of neurodegenerative diseases with Subtype and Stage Inference. \emph{Nat. Commun.} \textbf{9}, 4273 (2018).
\item Raj, A., Kuceyeski, A. \& Weiner, M. A network diffusion model of disease progression in dementia. \emph{Neuron} \textbf{73}, 1204--1215 (2012).
\item Vogel, J. W. et al. Four distinct trajectories of tau deposition identified in Alzheimer's disease. \emph{Nat. Med.} \textbf{27}, 871--881 (2021).
\item Vogel, J. W. et al. Connectome-based modelling of neurodegenerative diseases: towards precision medicine and mechanistic insight. \emph{Nat. Rev. Neurosci.} \textbf{24}, 620--639 (2023).
\item Young, A. L. et al. Data-driven modelling of neurodegenerative disease progression: thinking outside the black box. \emph{Nat. Rev. Neurosci.} \textbf{25}, 111--130 (2024).
\item Wang, J., Mao, Y., Liu, X. \& Hao, W. Learning patient-specific spatial biomarker dynamics via operator learning for Alzheimer's disease progression. \emph{npj Syst. Biol. Appl.} \textbf{12}, 104 (2026).
\item Abbott, N. J., Rönnbäck, L. \& Hansson, E. Astrocyte--endothelial interactions at the blood--brain barrier. \emph{Nat. Rev. Neurosci.} \textbf{7}, 41--53 (2006).
\item Sofroniew, M. V. \& Vinters, H. V. Astrocytes: biology and pathology. \emph{Acta Neuropathol.} \textbf{119}, 7--35 (2010).
\item Iliff, J. J. et al. A paravascular pathway facilitates CSF flow through the brain parenchyma and the clearance of interstitial solutes, including amyloid $\beta$. \emph{Sci. Transl. Med.} \textbf{4}, 147ra111 (2012).
\item Vainchtein, I. D. \& Molofsky, A. V. Astrocytes and microglia: in sickness and in health. \emph{Trends Neurosci.} \textbf{43}, 144--154 (2020).
\item Leng, F. \& Edison, P. Neuroinflammation and microglial activation in Alzheimer disease: where do we go from here? \emph{Nat. Rev. Neurol.} \textbf{17}, 157--172 (2021).
\item Deczkowska, A. et al. Disease-associated microglia: a universal immune sensor of neurodegeneration. \emph{Cell} \textbf{173}, 1073--1081 (2018).
\item Keren-Shaul, H. et al. A unique microglia type associated with restricting development of Alzheimer's disease. \emph{Cell} \textbf{169}, 1276--1290.e17 (2017).
\item Hansen, D. V., Hanson, J. E. \& Sheng, M. Microglia in Alzheimer's disease. \emph{J. Cell Biol.} \textbf{217}, 459--472 (2018).
\item Zlokovic, B. V. The blood--brain barrier in health and chronic neurodegenerative disorders. \emph{Neuron} \textbf{57}, 178--201 (2008).
\item Butterfield, D. A. \& Halliwell, B. Oxidative stress, dysfunctional glucose metabolism and Alzheimer disease. \emph{Nat. Rev. Neurosci.} \textbf{20}, 148--160 (2019).
\item Korte, N., Nortley, R. \& Attwell, D. Cerebral blood flow decrease as an early pathological mechanism in Alzheimer's disease. \emph{Acta Neuropathol.} \textbf{140}, 793--810 (2020).
\item Verkhratsky, A. \& Zorec, R. Neuroglia in cognitive reserve. \emph{Mol. Psychiatry} \textbf{29}, 3962--3967 (2024).
\item Hall, C. M., Moeendarbary, E. \& Sheridan, G. K. Mechanobiology of the brain in ageing and Alzheimer's disease. \emph{Eur. J. Neurosci.} \textbf{53}, 3851--3878 (2021).
\item Edison, P. Astroglial activation: current concepts and future directions. \emph{Alzheimers Dement.} \textbf{20}, 3034--3053 (2024).
\item de Sá Hayashide, L. et al. From neuron-centric to glia-centric: how aging glial networks drive neurodegenerative disease. \emph{J. Neurochem.} \textbf{170}, e70361 (2026).
\item Bice, P. J., Nho, K., Ertekin-Taner, N. \& Saykin, A. J. Systems biology of Alzheimer's disease: a scoping review of key pathways and mechanisms. \emph{Mol. Neurodegener.} \textbf{21}, 21 (2026).
\item Yang, Z. et al. Cell-type-specific Alzheimer's disease polygenic risk scores are associated with distinct disease processes in Alzheimer's disease. \emph{Nat. Commun.} \textbf{14}, 7659 (2023).
\item Corder, E. H. et al. Gene dose of apolipoprotein E type 4 allele and the risk of Alzheimer's disease in late onset families. \emph{Science} \textbf{261}, 921--923 (1993).
\item Kunkle, B. W. et al. Genetic meta-analysis of diagnosed Alzheimer's disease identifies new risk loci and implicates A$\beta$, tau, immunity and lipid processing. \emph{Nat. Genet.} \textbf{51}, 414--430 (2019).
\item Jansen, I. E. et al. Genome-wide meta-analysis identifies new loci and functional pathways influencing Alzheimer's disease risk. \emph{Nat. Genet.} \textbf{51}, 404--413 (2019).
\item Mathys, H. et al. Single-cell transcriptomic analysis of Alzheimer's disease. \emph{Nature} \textbf{570}, 332--337 (2019).
\item Grubman, A. et al. A single-cell atlas of entorhinal cortex from individuals with Alzheimer's disease reveals cell-type-specific gene expression regulation. \emph{Nat. Neurosci.} \textbf{22}, 2087--2097 (2019).
\item Stern, Y. Cognitive reserve in ageing and Alzheimer's disease. \emph{Lancet Neurol.} \textbf{11}, 1006--1012 (2012).
\item Barulli, D. \& Stern, Y. Efficiency, capacity, compensation, maintenance, plasticity: emerging concepts in cognitive reserve. \emph{Trends Cogn. Sci.} \textbf{17}, 502--509 (2013).
\item Ahmed, S., Haigh, A.-M. F., de Jager, C. A. \& Garrard, P. Connected speech as a marker of disease progression in autopsy-proven Alzheimer's disease. \emph{Brain} \textbf{136}, 3727--3737 (2013).
\item Hajjar, I. et al. Development of digital voice biomarkers and associations with cognition, cerebrospinal biomarkers and neural representation in early Alzheimer's disease. \emph{Alzheimers Dement. Amst.} \textbf{15}, e12393 (2023).
\item Ostoja-Starzewski, M. \emph{Microstructural Randomness and Scaling in Mechanics of Materials} (CRC Press, 2008).
\item Iturria-Medina, Y., Carbonell, F. M. \& Evans, A. C. Multimodal imaging-based therapeutic fingerprints for optimizing personalized interventions: application to neurodegeneration. \emph{Neuroimage} \textbf{179}, 40--50 (2018).
\item Borchert, R. J. et al. Artificial intelligence for diagnostic and prognostic neuroimaging in dementia: a systematic review. \emph{Alzheimers Dement.} \textbf{19}, 5885--5904 (2023).
\item Qiu, S. et al. Multimodal deep learning for Alzheimer's disease dementia assessment. \emph{Nat. Commun.} \textbf{13}, 3404 (2022).
\item Xue, C. et al. AI-based differential diagnosis of dementia etiologies on multimodal data. \emph{Nat. Med.} \textbf{30}, 2977--2989 (2024).
\item Jasodanand, V. H. et al. AI-driven fusion of multimodal data for Alzheimer's disease biomarker assessment. \emph{Nat. Commun.} \textbf{16}, 7407 (2025).
\item Qi, W. et al. Alzheimer's disease digital biomarkers multidimensional landscape and AI model scoping review. \emph{npj Digit. Med.} \textbf{8}, 366 (2025).
\item Wiechmann, D. et al. Detecting CSF-validated Alzheimer's disease from spontaneous speech in German: an interpretable end-to-end machine-learning framework. \emph{Front. Neurol.} \textbf{17}, 1780783 (2026).
\item Azadmaleki, H. et al. SpeechCARE: dynamic multimodal modeling for cognitive screening in diverse linguistic and speech task contexts. \emph{npj Digit. Med.} \textbf{8}, 677 (2025).
\item Shankar, R. et al. A systematic review of explainable artificial intelligence methods for speech-based cognitive decline detection. \emph{npj Digit. Med.} \textbf{8}, 724 (2025).
\item Polk, S. E. et al. A scoping review of remote and unsupervised digital cognitive assessments in preclinical Alzheimer's disease. \emph{npj Digit. Med.} \textbf{8}, 266 (2025).
\item Matias, I. et al. Passive digital health technologies for Alzheimer's disease screening and diagnosis: a systematic review. \emph{npj Digit. Med.} \textbf{9}, 496 (2026).
\item Javeed, A. et al. Machine learning for dementia prediction: a systematic review and future research directions. \emph{J. Med. Syst.} \textbf{47}, 17 (2023).
\item Yousefi, M. et al. Machine learning based algorithms for virtual early detection and screening of neurodegenerative and neurocognitive disorders: a systematic review. \emph{Front. Neurol.} \textbf{15}, 1413071 (2024).
\item Veronese, N. et al. Clinical prediction models using artificial intelligence approaches in dementia. \emph{Aging Clin. Exp. Res.} \textbf{37}, 233 (2025).
\item Vaghari, D. et al. AI-guided patient stratification improves outcomes and efficiency in the AMARANTH Alzheimer's Disease clinical trial. \emph{Nat. Commun.} \textbf{16}, 6244 (2025).
\item Bracher-Smith, M. et al. Machine learning in Alzheimer's disease genetics. \emph{Nat. Commun.} \textbf{16}, 6726 (2025).
\item Zhang, Z., Zou, Z., Kuhl, E. \& Karniadakis, G. E. Discovering a reaction--diffusion model for Alzheimer's disease by combining physics-informed neural networks with symbolic regression. \emph{Comput. Methods Appl. Mech. Eng.} \textbf{419}, 116647 (2024).
\item Raissi, M., Perdikaris, P. \& Karniadakis, G. E. Physics-informed neural networks: a deep learning framework for solving forward and inverse problems involving nonlinear partial differential equations. \emph{J. Comput. Phys.} \textbf{378}, 686--707 (2019).
\item Karniadakis, G. E. et al. Physics-informed machine learning. \emph{Nat. Rev. Phys.} \textbf{3}, 422--440 (2021).
\item National Academies of Sciences, Engineering, and Medicine. \emph{Foundational Research Gaps and Future Directions for Digital Twins} (National Academies Press, 2024).
\item Wright, L. \& Davidson, S. How to tell the difference between a model and a digital twin. \emph{Adv. Model. Simul. Eng. Sci.} \textbf{7}, 13 (2020).
\item Sanz Perl, Y. et al. Perturbations in dynamical models of whole-brain activity dissociate between the level and stability of consciousness. \emph{Nat. Commun.} \textbf{14}, 4129 (2023).
\item Oberkampf, W. L. \& Roy, C. J. \emph{Verification and Validation in Scientific Computing} (Cambridge Univ. Press, 2010).
\item Kennedy, M. C. \& O'Hagan, A. Bayesian calibration of computer models. \emph{J. R. Stat. Soc. Ser. B Stat. Methodol.} \textbf{63}, 425--464 (2001).
\item Tarantola, A. \emph{Inverse Problem Theory and Methods for Model Parameter Estimation} (SIAM, 2005).
\item Goldsack, J. C. et al. Verification, analytical validation, and clinical validation (V3): the foundation of determining fit-for-purpose for biometric monitoring technologies. \emph{npj Digit. \textbf{Med.}}3**, 55 (2020).
\item Wilkinson, M. D. et al. The FAIR Guiding Principles for scientific data management and stewardship. \emph{Sci. Data} \textbf{3}, 160018 (2016).
\item Gorgolewski, K. J. et al. The Brain Imaging Data Structure, a format for organizing and describing outputs of neuroimaging experiments. \emph{Sci. Data} \textbf{3}, 160044 (2016).
\item ASME. \emph{Verification, Validation, and Uncertainty Quantification Terminology in Computational Modeling and Simulation ASME VVUQ 1-2022} (American Society of Mechanical Engineers, 2022).
\item ASME. \emph{Assessing Credibility of Computational Modeling Through Verification and Validation: Application to Medical Devices ASME V\textbf{\&} V 40-2018} (American Society of Mechanical Engineers, 2018).
\item U.S. Food and Drug Administration. \emph{Assessing the Credibility of Computational Modeling and Simulation in Medical Device Submissions: Guidance for Industry and Food and Drug Administration Staff} (FDA, 2023).
\item Petersen, R. C. et al. Alzheimer's Disease Neuroimaging Initiative (ADNI): clinical characterization. \emph{Neurology} \textbf{74}, 201--209 (2010).
\item Weiner, M. W. et al. The Alzheimer's Disease Neuroimaging Initiative: a review of papers published since its inception. \emph{Alzheimers Dement.} \textbf{9}, e111--e194 (2013).
\item LaMontagne, P. J. et al. OASIS-3: longitudinal neuroimaging, clinical, and cognitive dataset for normal aging and Alzheimer disease. \emph{medRxiv} \url{https://doi.org/10.1101/2019.12.13.19014902} (2019).
\item Solomon, A. et al. European Prevention of Alzheimer's Dementia Longitudinal Cohort Study (EPAD LCS): study protocol. \emph{BMJ Open} \textbf{8}, e021017 (2018).
\item Poewe, W. et al. Parkinson disease. \emph{Nat. Rev. Dis. Primers} \textbf{3}, 17013 (2017).
\item Braak, H. et al. Staging of brain pathology related to sporadic Parkinson's disease. \emph{Neurobiol. Aging} \textbf{24}, 197--211 (2003).
\item Jucker, M. \& Walker, L. C. Self-propagation of pathogenic protein aggregates in neurodegenerative diseases. \emph{Nature} \textbf{501}, 45--51 (2013).
\end{refs}
\end{document}


\thispagestyle{plain}

\begin{flushleft}
{\footnotesize\sffamily\bfseries\color{npjblue}SUPPLEMENTARY INFORMATION}\\[2pt]
{\color{rulegrey!60}\rule{\textwidth}{0.5pt}}\\[10pt]
{\bfseries\LARGE Towards a physics-informed multiscale digital twin for precision medicine in Alzheimer's disease\par}
\vspace{10pt}
{\normalsize Aida Nonn\textsuperscript{1,\Letter}, Elma Kerz\textsuperscript{2}, Daniel Wiechmann\textsuperscript{2,3} \& Paul Edison\textsuperscript{4}\par}
\vspace{7pt}
{\footnotesize
\textsuperscript{1}Ostbayerische Technische Hochschule Regensburg, Regensburg, Germany.
\textsuperscript{2}Exaia Technologies GmbH, Aachen, Germany.
\textsuperscript{3}University of Amsterdam, Amsterdam, The Netherlands.
\textsuperscript{4}Imperial College London, London, UK.\\[2pt]
\textsuperscript{\Letter}Correspondence: Aida Nonn (\href{mailto:aida.nonn@oth-regensburg.de}{aida.nonn@oth-regensburg.de}).\par}
\end{flushleft}
\vspace{6pt}
{\footnotesize\sffamily This file contains Supplementary Notes 1--4, Supplementary Tables 1--3 and Supplementary References.\par}
\vspace{4pt}

\section*{Supplementary Note 1 $|$ State variables, observation layers and cross-scale parameter passing}

This note expands the state-variable and observation architecture described in the main text. This glial-centred realisation of PIM-BrainTwin represents Alzheimer's disease as a coupled multiscale dynamical system in which biological support states, pathology burden and functional-network physiology evolve together.\sfcite{1--3} The supplementary mapping makes the architecture auditable by linking each biological domain to a latent state, measurement class, observation equation and role in participant-specific state estimation.

The latent system-state vector is

\begin{equation*} X_t=\left(A_t,\,V_t,\,I_t,\,M_t,\,P_t,\,F_t\right)^{\!\top}, \end{equation*}

where $A_t$ denotes glial adaptive capacity, $V_t$ vascular--haemodynamic support, $I_t$ inflammatory--clearance state, $M_t$ neuronal--metabolic support, $P_t$ amyloid--tau burden and $F_t$ functional-network physiology. Cognitive, speech and behavioural performance are represented separately in the observation layer, denoted $C_t$, because they express the functional consequences of biological state rather than constituting a primary latent physiological state.

In this glial-centred realisation, $A_t$ is the focal latent capacity variable used to operationalise the safety margin, but it is not synonymous with system-level compensatory capacity. System-level compensatory capacity emerges from the coupled configuration of $A_t$, $V_t$, $I_t$, $M_t$, $P_t$ and $F_t$ under the prevailing biological load. The scalar safety margin $S_t = A_t - A_t^{\mathrm{crit}}$ is therefore a model-derived summary of compensatory stability within this specific realisation, with $A_t^{\mathrm{crit}}$ allowed to depend on the state of the interacting subsystems.

This separation is important for interpreting cognitive reserve. A participant can show relatively preserved speech or cognitive performance despite altered network physiology. In the model, such divergence is informative: it constrains the observation function linking $F_t$ to measured cognition and behaviour, rather than forcing cognition to define the latent physiological state directly. Neuroglial contributions to cognitive reserve can be represented through the influence of $A_t$ and its coupling to $F_t$ on this observation function, rather than by introducing cognitive reserve as an additional primary state variable.\sfcite{4}

The generic state equation is written as

\begin{equation*} \frac{\mathrm{d}X_t}{\mathrm{d}t} = f\!\left(X_t,\,u_t,\,p(x,t);\,\theta\right), \end{equation*}

where $u_t$ denotes perturbations or interventions, $p(x,t)$ denotes spatially distributed amyloid--tau pathology and $\theta$ denotes participant- or population-level parameters. Here, $P_t$ denotes the participant-level pathology state, whereas $p(x,t)$ denotes its spatially resolved field representation for propagation and tissue-scale simulation. The function $f$ encodes depletion, repair, propagation and feedback terms. A given biological hypothesis becomes part of the model only when it can be expressed through one or more of these components: a state variable, an interaction term, a parameter prior, an observation equation or an intervention handle.

At the micro scale, cellular and neurovascular-unit observations estimate repair kinetics, depletion kinetics, genotype-dependent vulnerability and persistent loss of recoverable capacity. These measurements constrain parameters governing glial adaptive capacity, inflammatory--clearance dynamics, vascular support and metabolic repair gating.\sfcite{5--15}

At the meso scale, multiscale computational homogenisation, implemented through statistical-volume-element (SVE) representations where appropriate, translates microstructural and biomechanical properties into effective tissue coefficients. Relevant quantities include endfoot geometry, aquaporin-4 polarity, perivascular architecture, basement-membrane properties, vascular compliance, tissue transport, clearance, porosity, stiffness and haemodynamic loading. These effective coefficients define the physical parameter-passing layer between cellular systems and participant-level brain dynamics.\sfcite{10--16}

At the macro scale, participant-specific observations update the latent trajectory. Imaging, fluid biomarkers, transcriptomics, electrophysiology, speech and cognition do not define the state variables directly. They enter through observation equations that link each measurement to one or more latent states, with biological cross-loading represented explicitly. For example, GFAP may preferentially constrain astrocyte-centred adaptive capacity, while also loading on inflammatory state; FDG-PET constrains neuronal--metabolic support but can also reflect downstream network dysfunction; and speech constrains functional expression while requiring an observation model that accounts for reserve, task demands and measurement conditions.\sfcite{5--9,12,14}

\begin{landscape}
\begin{footnotesize}
\setlength{\tabcolsep}{4pt}\renewcommand{\arraystretch}{1.25}
\begin{longtable}{>{\raggedright\arraybackslash}p{0.100\linewidth}>{\raggedright\arraybackslash}p{0.080\linewidth}>{\raggedright\arraybackslash}p{0.220\linewidth}>{\raggedright\arraybackslash}p{0.210\linewidth}>{\raggedright\arraybackslash}p{0.210\linewidth}>{\raggedright\arraybackslash}p{0.080\linewidth}}
\caption*{\textbf{Supplementary Table 1 $|$ Biological domains, state variables, observations and observation roles}}\\
\toprule
\textbf{Biological domain} & \textbf{State variable or layer} & \textbf{Biological meaning} & \textbf{Candidate observations} & \textbf{Observation role} & \textbf{Typical temporal resolution} \\
\midrule\endfirsthead
\multicolumn{6}{l}{\footnotesize\emph{Supplementary Table 1 (continued)}}\\[2pt]
\toprule
\textbf{Biological domain} & \textbf{State variable or layer} & \textbf{Biological meaning} & \textbf{Candidate observations} & \textbf{Observation role} & \textbf{Typical temporal resolution} \\
\midrule\endhead
\bottomrule\endfoot
Glial adaptive capacity & $A_t$ & Stress--recovery capacity of astrocyte-centred support systems, including neurovascular-unit support, metabolic buffering, ionic and synaptic homeostasis and clearance-related functions & Astrocyte-reactivity PET where available, GFAP, astrocyte-enriched transcriptomic signatures, iPSC-derived astrocyte and neurovascular-unit perturbation readouts & Constrains adaptive capacity, repair kinetics, depletion rates and persistent loss of recoverable capacity & Low to intermediate \\ \addlinespace[2pt]
Vascular--haemodynamic support & $V_t$ & Perfusion, vascular reactivity, endothelial support, BBB integrity, vascular compliance and haemodynamic loading & ASL-MRI, cerebrovascular reactivity, BBB markers, vascular-risk measures, DTI-derived vascular or tissue context where available & Constrains vascular support and vascular-to-inflammatory or vascular-to-metabolic coupling & Low to intermediate \\ \addlinespace[2pt]
Inflammatory--clearance state & $I_t$ & Microglial activation, astrocyte--microglia signalling, immune resolution and inflammatory components of clearance & TSPO-PET, sTREM2, cytokines, inflammatory proteomics, microglia-enriched transcriptomic signatures & Constrains inflammatory gain, resolution, clearance coupling and inflammatory depletion of adaptive capacity & Low to intermediate \\ \addlinespace[2pt]
Neuronal--metabolic support & $M_t$ & Glucose metabolism, mitochondrial function, energetic support and repair gating & FDG-PET, metabolic markers, mitochondrial readouts, oxidative-stress markers, metabolic transcriptomic signatures & Constrains repair gating, metabolic vulnerability and fatigue-like depletion & Low \\ \addlinespace[2pt]
Amyloid--tau burden & $P_t$ & Amyloid and tau burden, regional pathology accumulation, spatial propagation and clearance & Amyloid PET, tau PET, CSF or plasma A$\beta$, p-tau and related pathology markers & Constrains pathology burden, propagation field, clearance terms and pathology-to-capacity coupling & Low \\ \addlinespace[2pt]
Functional-network physiology & $F_t$ & Electrophysiological and functional-network state, including network stability, synchrony and physiological compensation & EEG, MEG where available, resting-state or task fMRI, connectivity measures & Constrains network physiology and transition from biological pathology to functional expression & Intermediate \\ \addlinespace[2pt]
Functional expression & $C_t$ & Reserve-modulated cognition, speech and behaviour & Cognitive testing, speech-derived digital biomarkers, behavioural measures, real-world functioning & Observation layer mapping $F_t$ and reserve-related parameters to measured performance & Intermediate to high \\ \addlinespace[2pt]
Genetic susceptibility & Parameter-prior layer & Inherited susceptibility affecting baseline compensatory capacity, repair, depletion-channel weights and interaction parameters & APOE genotype, genome-wide AD PRS, cell-type-weighted PRS, rare-variant burden where available & Provides structured priors on parameters rather than direct state measurements & Time-invariant \\ \addlinespace[2pt]
Cell-state molecular readouts & Observation and prior layer & Dynamic molecular state of glial, neuronal, vascular and immune cell populations & Single-cell, single-nucleus, spatial transcriptomics, proteomics & Constrains observation maps and state-specific molecular priors & Low \\ \addlinespace[2pt]
Intervention or perturbation & $u_t$ & Therapeutic, behavioural, physiological or experimental perturbation & Treatment exposure, neuromodulation, vascular challenge, anti-inflammatory intervention, metabolic intervention & Defines counterfactual intervention scenarios and perturbation-response tests & Context-dependent \\ \addlinespace[2pt]
\end{longtable}
\end{footnotesize}
\end{landscape}

\clearpage
\section*{Supplementary Note 2 $|$ Genetic susceptibility and priors on repair--depletion dynamics}

This note expands the main-text statement that genetic susceptibility enters the architecture through hierarchical priors on baseline compensatory capacity, repair--depletion kinetics and subsystem-specific interaction strengths. Genetic susceptibility is represented as a structured prior architecture rather than as a single disease-risk covariate. Across brain disorders, inherited liability spans rare high-effect variants, copy-number variation, rare coding variation, APOE-related risk and highly polygenic configurations of common alleles.\sfcite{17--22} Rare variants and large-effect mutations can strongly shape familial or early-onset disease mechanisms, whereas common late-onset Alzheimer's disease risk is distributed across APOE, immune, lipid, endosomal, vascular and cell-type-specific genetic pathways.\sfcite{23--27}

These genetic effects are not uniformly expressed across the brain or lifespan. Functional genomic studies show that disease-associated genetic signals are instantiated through cell-type-specific regulatory landscapes and spatiotemporal windows of vulnerability.\sfcite{5--9,20,27} In this realisation, the principle is applied conservatively: genetic susceptibility enters as a prior on vulnerability and response parameters, not as a direct measurement of the latent system state.

Inherited DNA variation is time-invariant, whereas glial adaptive capacity, inflammatory--clearance state, vascular support, metabolic support and functional-network physiology evolve over time. Genetic information therefore enters the model upstream of the latent trajectory, shaping prior distributions over parameters that determine how an individual system responds to pathological, inflammatory, metabolic, vascular and mechanical load.

Let $z_i^{\mathrm{gen}}$ denote the genetic-susceptibility vector for participant i. This vector may include APOE genotype or $\varepsilon_{4}$ dosage, genome-wide Alzheimer's disease polygenic risk, cell-type-weighted polygenic risk scores and, where available, rare-variant burden in genes implicated in immune, lipid, endosomal, vascular or clearance pathways.\sfcite{24--32} In familial or early-onset Alzheimer's disease contexts, the vector may also include pathogenic variants in APP, PSEN1 or PSEN2, which are treated as strong pathway-specific priors on amyloid-centred initiation rather than as generic late-onset depletion parameters.\sfcite{33--35} These quantities are measured once and used as hierarchical covariates on participant-specific parameters. They do not define $A_t$, $I_t$, $V_t$, $M_t$, $P_t$ or $F_t$ directly.

The core adaptive-capacity dynamics are written as a repair--depletion balance,

\begin{equation*} \tau_A \frac{\partial A}{\partial t} = \gamma_0\,\sigma_\gamma(M,V)\left[\left(1-D_{\mathrm{irr}}\right)-A\right] - \delta_A A. \end{equation*}

Here, $A$ is glial adaptive capacity, $\gamma_0$ is the maximal repair rate, $\sigma_\gamma(M,V)$ gates repair through metabolic and vascular support, $D_{\mathrm{irr}}$ denotes persistent loss of recoverable capacity and $\delta_A$ is the instantaneous depletion rate. The depletion rate is decomposed into biologically interpretable channels,

\begin{equation*} \delta_A = a_I\,I + a_C\left(1-\phi_A\right) + a_H\,H. \end{equation*}

Here, $a_I$ weights inflammatory depletion, $a_C$ weights clearance/endfoot depletion and $a_H$ weights haemodynamic or mechanical-load depletion. $H$ is a relaxing load-memory state, allowing transient haemodynamic load to be separated from persistent injury. Persistent capacity loss evolves separately as

\begin{equation*} \frac{\partial D_{\mathrm{irr}}}{\partial t} = \left(1-D_{\mathrm{irr}}\right)\left[\,d_S\,\langle -S\rangle_{+} + d_I\,\langle I-I_0\rangle_{+} + d_H\,\langle H-H_0\rangle_{+}\right]. \end{equation*}

This separation prevents reversible stress, current depletion and irreversible loss of recoverable capacity from being conflated. It also allows genetic susceptibility to influence different components of vulnerability in a biologically specific way.

Genetic susceptibility enters through hierarchical priors on baseline compensatory capacity, repair kinetics, depletion-channel weights and persistent-injury thresholds. A generic implementation is

\begin{equation*} \operatorname{logit}\!\left(A_{0,i}\right) \sim \mathcal{N}\!\left(\mu_{A_0} + \lambda^{A_0}_{\mathrm{APOE}}\,\mathrm{APOE4}_i + \lambda^{A_0}_{\mathrm{Ast}}\,\mathrm{PRS}^{\mathrm{Ast}}_i + \lambda^{A_0}_{\mathrm{Vasc}}\,\mathrm{PRS}^{\mathrm{Vasc}}_i,\; \sigma^2_{A_0}\right), \end{equation*}

\begin{equation*} \log \gamma_{0,i} \sim \mathcal{N}\!\left(\mu_{\gamma} + \lambda^{\gamma}_{\mathrm{Ast}}\,\mathrm{PRS}^{\mathrm{Ast}}_i + \lambda^{\gamma}_{\mathrm{Met}}\,\mathrm{PRS}^{\mathrm{Met}}_i,\; \sigma^2_{\gamma}\right), \end{equation*}

\begin{equation*} \log a_{I,i} \sim \mathcal{N}\!\left(\mu_{a_I} + \lambda^{a_I}_{\mathrm{Mic}}\,\mathrm{PRS}^{\mathrm{Mic}}_i + \lambda^{a_I}_{\mathrm{APOE}}\,\mathrm{APOE4}_i,\; \sigma^2_{a_I}\right), \end{equation*}

\begin{equation*} \log a_{C,i} \sim \mathcal{N}\!\left(\mu_{a_C} + \lambda^{a_C}_{\mathrm{Ast}}\,\mathrm{PRS}^{\mathrm{Ast}}_i + \lambda^{a_C}_{\mathrm{APOE}}\,\mathrm{APOE4}_i + \lambda^{a_C}_{\mathrm{Clr}}\,\mathrm{PRS}^{\mathrm{Clr}}_i,\; \sigma^2_{a_C}\right), \end{equation*}

\begin{equation*} \log a_{H,i} \sim \mathcal{N}\!\left(\mu_{a_H} + \lambda^{a_H}_{\mathrm{Vasc}}\,\mathrm{PRS}^{\mathrm{Vasc}}_i + \lambda^{a_H}_{\mathrm{Ast}}\,\mathrm{PRS}^{\mathrm{Ast}}_i,\; \sigma^2_{a_H}\right). \end{equation*}

For familial or early-onset Alzheimer's disease settings, an additional pathology-forcing prior may be specified as

\begin{equation*} \log P_{0,i} \sim \mathcal{N}\!\left(\mu_{P_0} + \lambda^{P_0}_{\mathrm{APP}}\,\mathrm{APP}_i + \lambda^{P_0}_{\mathrm{PSEN1}}\,\mathrm{PSEN1}_i + \lambda^{P_0}_{\mathrm{PSEN2}}\,\mathrm{PSEN2}_i,\; \sigma^2_{P_0}\right), \end{equation*}

or through an equivalent pathway-specific prior on amyloid production, amyloid clearance or pathology--capacity coupling. This term is not intended for generic late-onset Alzheimer's disease unless familial or early-onset genetic evidence is present.

Here,

\begin{equation*} \mathrm{PRS}^{\mathrm{Ast}},\; \mathrm{PRS}^{\mathrm{Mic}},\; \mathrm{PRS}^{\mathrm{Vasc}},\; \mathrm{PRS}^{\mathrm{Met}},\; \mathrm{PRS}^{\mathrm{Clr}} \end{equation*}

denote astrocyte-, microglia-, vascular-, metabolic- and clearance-weighted genetic-risk scores. The coefficients $\lambda$ encode how strongly genetic susceptibility shifts each prior distribution. These effects are probabilistic and regularised; they are not deterministic claims that genetic risk causes a fixed trajectory. If posterior checks do not support participant-level resolution, the corresponding genetic effect remains a population- or subgroup-level prior.

Cell-type-weighted genetic scores are treated as hypothesis-generating priors, not as direct mechanistic proof.\sfcite{27} Their biological interpretation depends on the quality of the underlying genetic association, the specificity of the cell-type annotation, the disease stage being modelled and the extent to which the resulting prior improves calibration or prediction. The model therefore does not assume that a genetic signal maps one-to-one onto a single cell type or biological mechanism.

To avoid treating genetic susceptibility as uniformly expressed across all brain regions and disease phases, the model may include external spatiotemporal weighting functions. These weights can be derived from single-cell, single-nucleus or regulatory genomic atlases and used to modulate whether a given genetic signal is most plausibly interpreted through astrocytic, microglial, endothelial, neuronal, oligodendrocytic, metabolic or clearance-related pathways. In the Alzheimer's disease implementation, late-life ageing, glial state, vascular integrity, lipid handling, endosomal trafficking, inflammation, clearance and metabolic support are prioritised over developmental regulatory windows unless disease-specific evidence justifies otherwise.\sfcite{5--9,20,27}

The expected directional assumptions are as follows. APOE $\varepsilon_{4}$ dosage and high genome-wide Alzheimer's disease risk may lower the prior mean of baseline adaptive capacity, reduce repair efficiency, increase clearance-related depletion and increase pathology--capacity coupling. Astrocyte-weighted risk may shift priors on $A_0$, $\gamma_0$, $a_C$, $a_H$ and persistent injury, reflecting vulnerability in astrocyte-centred support, endfoot organisation, metabolic coupling and recovery. Microglia-weighted risk primarily shifts priors on inflammatory activation and resolution parameters, including $a_I$, inflammatory gain and clearance-state dynamics. Vascular-weighted risk shifts priors on vascular support, vascular-to-inflammatory coupling and haemodynamic-load depletion. Metabolic-weighted risk shifts priors on repair gating and neuronal--metabolic support. Rare variants, where available, can be represented as additional covariates or subgroup priors, but only if sample size and variant frequency support interpretable inference.\sfcite{23--27}

Rare high-penetrance AD variants are handled separately from polygenic late-onset risk. APP, PSEN1 and PSEN2 variants, where present, are expected to modify pathology initiation or amyloid-centred forcing more strongly than generic glial depletion parameters. Rare coding variants in TREM2, SORL1, ABCA7, PLCG2, ABI3 and related immune, lipid, endosomal or clearance genes are treated as pathway-specific susceptibility priors, usually at subgroup level unless sequencing depth, sample size and variant frequency support more precise inference.\sfcite{28--32}

These assumptions make genetic susceptibility testable within the model. High genetic risk should not merely predict higher disease probability; it should improve inference over specific dynamical quantities if the assumed biological mapping is correct. For example, astrocyte-weighted genetic risk should improve prediction of lower baseline $A_0$, slower recovery, greater clearance/endfoot depletion or greater persistent loss of recoverable capacity. Microglia-weighted risk should improve prediction of inflammatory gain, slower resolution or stronger inflammatory depletion. Vascular-weighted risk should improve prediction of vascular support, haemodynamic-load sensitivity or vascular-to-inflammatory coupling. If these mappings fail posterior predictive checks, the genetic prior structure should be weakened, revised or removed.

The quantitative impact of genetic susceptibility can be estimated in large retrospective cohorts, such as ADNI, by measuring how APOE genotype, genome-wide PRS and cell-type-weighted PRS shift longitudinal biomarker trajectories and inferred depletion-related parameters. The relevant estimands include posterior shifts in $A_0$, $\gamma_0$, $a_I$, $a_C$, $a_H$, $D_{\mathrm{irr}}$, $k^{A}_{\mathrm{clr}}$, pathology--capacity coupling and the safety-margin trajectory $S_t$. Genetic effects should be evaluated by held-out prediction, posterior calibration, precision gain and ablation tests comparing models with and without genetic-prior terms.

Prior information from genetic susceptibility is combined with experimental and molecular evidence. Human iPSC-derived astrocyte and neurovascular-unit experiments estimate stress--recovery kinetics, depletion slopes and persistent loss of recoverable capacity under inflammatory, metabolic, amyloid--tau, clearance and mechanical or haemodynamic stressors. Transcriptomic cell-state signatures provide dynamic molecular observations of current cellular state, whereas genetic susceptibility provides time-invariant priors on vulnerability and response parameters. This preserves the distinction between inherited liability, molecular state and longitudinal dynamics.

This triangulation is essential for causal interpretation. Large cohorts can estimate quantitative associations and calibrate useful priors, but they cannot by themselves prove that genetic susceptibility causally alters a specific depletion coefficient. For stronger interpretation, cohort-derived estimates must be triangulated with iPSC-derived astrocyte and neurovascular-unit perturbation experiments, transcriptomic cell-state evidence, imaging readouts and prospective validation.

The model therefore interprets genetic susceptibility as a parameter-shaping layer. It influences where the trajectory starts, how rapidly adaptive capacity is depleted under load, how effectively the system repairs after perturbation and how easily transient stress becomes persistent injury. It does not replace multimodal observation, and it does not define the safety margin by itself. At participant level, the safety margin is represented as $S_t = A_t - A_t^{\mathrm{crit}}$. Spatially resolved implementations generalise this quantity to

\begin{equation*} S(x,t) = A(x,t) - A^{\mathrm{crit}}(x,t), \end{equation*}

where $A^{\mathrm{crit}}$ may vary with prevailing biological load and the state of interacting vascular, inflammatory, metabolic and pathological processes. Genetic information contributes through prior distributions over the parameters that shape $A(x,t)$, $A^{\mathrm{crit}}(x,t)$ and posterior precision.

In summary, genetics explains inherited susceptibility, whereas this glial-centred realisation of PIM-BrainTwin examines how that susceptibility is expressed through time-varying physiological trajectories. Genetic information is therefore foundational but not sufficient: it constrains vulnerability, repair and depletion priors, while longitudinal multimodal observations determine how the coupled system evolves.

\begin{landscape}
\begin{footnotesize}
\setlength{\tabcolsep}{4pt}\renewcommand{\arraystretch}{1.25}
\begin{longtable}{>{\raggedright\arraybackslash}p{0.130\linewidth}>{\raggedright\arraybackslash}p{0.190\linewidth}>{\raggedright\arraybackslash}p{0.170\linewidth}>{\raggedright\arraybackslash}p{0.170\linewidth}>{\raggedright\arraybackslash}p{0.250\linewidth}}
\caption*{\textbf{Supplementary Table 2 $|$ Genetic susceptibility metrics and prior roles in the glial-centred PIM-BrainTwin realisation}}\\
\toprule
\textbf{Genetic layer} & \textbf{Candidate metric} & \textbf{Expected model role} & \textbf{Candidate prior target} & \textbf{Interpretation and safeguards} \\
\midrule\endfirsthead
\multicolumn{5}{l}{\footnotesize\emph{Supplementary Table 2 (continued)}}\\[2pt]
\toprule
\textbf{Genetic layer} & \textbf{Candidate metric} & \textbf{Expected model role} & \textbf{Candidate prior target} & \textbf{Interpretation and safeguards} \\
\midrule\endhead
\bottomrule\endfoot
Rare high-penetrance AD variants & Pathogenic variants in APP, PSEN1 and PSEN2; familial AD status where relevant & Strong pathway-specific prior on amyloid-centred initiation and pathology forcing & $P_0$, amyloid production or clearance parameters, pathology--capacity coupling & Relevant mainly to familial or early-onset Alzheimer's disease. These variants are not treated as generic late-onset depletion parameters. \\ \addlinespace[2pt]
APOE genotype & $\varepsilon_{2}$/$\varepsilon_{3}$/$\varepsilon_{4}$ genotype; $\varepsilon_{4}$ dosage coded as 0, 1 or 2 & Broad susceptibility modifier affecting lipid handling, amyloid--tau handling, clearance, vascular integrity and glial response & $A_0$, $\gamma_0$, $a_C$, $a_I$, $k^{A}_{\mathrm{clr}}$, pathology--capacity coupling & APOE is treated as a structured prior on vulnerability and response kinetics, not as a direct disease-state variable. \\ \addlinespace[2pt]
Rare coding and pathway variants & Rare damaging variants or burden scores in TREM2, SORL1, ABCA7, PLCG2, ABI3 and related immune, lipid, endosomal or clearance genes & Pathway-specific susceptibility & $a_I$, $a_C$, inflammatory resolution, clearance efficiency, pathology--capacity coupling & Used as subgroup or pathway priors only when sequencing, sample size and variant frequency support interpretable inference. \\ \addlinespace[2pt]
Genome-wide AD PRS & Standardised Alzheimer's disease polygenic risk score & Global inherited liability across distributed common-variant architecture & $A_0$, depletion-channel weights, pathology--capacity coupling & Captures distributed genetic susceptibility across many loci. It is regularised and retained only if it improves calibration or prediction. \\ \addlinespace[2pt]
Astrocyte-weighted PRS & PRS weighted by astrocyte-enriched gene annotations or cell-type-specific genetic associations & Astrocyte-centred susceptibility & $A_0$, $\gamma_0$, $a_C$, $a_H$, $D_{\mathrm{irr}}$ & Higher astrocyte-weighted risk is hypothesised to lower baseline glial adaptive capacity, impair repair or increase persistent loss of recoverable capacity under stress. \\ \addlinespace[2pt]
Microglia-weighted PRS & PRS weighted by microglial gene annotations or immune-risk enrichment & Inflammatory susceptibility & $a_I$, inflammatory gain, inflammatory resolution rate, $I_t$ dynamics & Higher microglial risk is hypothesised to increase inflammatory activation, slow resolution or increase inflammatory depletion. \\ \addlinespace[2pt]
Vascular/endothelial PRS & Vascular, endothelial, pericyte or BBB-associated genetic weighting & Vascular and barrier vulnerability & $V_0$, vascular-to-inflammatory coupling, $a_H$, BBB-related observation priors & Higher vascular risk is hypothesised to reduce vascular support and increase haemodynamic or BBB-related depletion. \\ \addlinespace[2pt]
Metabolic/ mitochondrial PRS & Metabolic, mitochondrial or energetic pathway weighting & Repair and metabolic support & $\gamma_0$, $\sigma_\gamma(M,V)$, $M_0$, inflammatory-to-metabolic coupling & Higher metabolic risk is hypothesised to reduce repair gating and increase vulnerability to metabolic strain. \\ \addlinespace[2pt]
Clearance/endfoot-related risk & AQP4, endfoot, extracellular-matrix, lipid/endosomal or clearance-related pathway scores & Perivascular clearance and transport vulnerability & $a_C$, $\phi_A$, $k^{A}_{\mathrm{clr}}$, $D_{\mathrm{eff}}$ & Higher clearance-related risk is hypothesised to increase depletion through impaired endfoot/AQP4-dependent clearance. \\ \addlinespace[2pt]
Spatiotemporal regulatory context & External cell-type-, region- and ageing-stage annotations from single-cell, single-nucleus and regulatory genomic atlases & Weighting of PRS and pathway priors by cell type, region and disease phase & Cell-type-weighted priors on $A_0$, $a_I$, $a_C$, $V_0$, $M_0$ and coupling strengths & Prevents genetic risk from being treated as uniformly expressed across all cells, regions and disease stages. \\ \addlinespace[2pt]
Cross-disorder genetic architecture & Rare-variant, CNV and polygenic findings from neurodevelopmental and psychiatric genetics & Conceptual support for a multilevel prior architecture & Not directly used in Alzheimer's participant inference unless disease-specific evidence exists & Supports the general principle that genetic liability spans rare and common variation and is organised by cell type and timing. It should not be used as an Alzheimer's-specific parameter source without AD-specific validation. \\ \addlinespace[2pt]
Quantitative calibration in large cohorts & APOE, genome-wide PRS and cell-type-weighted PRS linked to longitudinal biomarkers in ADNI-like cohorts & Population- or subgroup-level estimation of genetic effect sizes & Posterior shifts in $A_0$, $\gamma_0$, $a_I$, $a_C$, $a_H$, $D_{\mathrm{irr}}$, $k^{A}_{\mathrm{clr}}$, pathology--capacity coupling and $S_t$ & Effects should be evaluated by held-out prediction, posterior calibration, precision gain and ablation tests comparing models with and without genetic-prior terms. \\ \addlinespace[2pt]
Experimental and molecular triangulation & iPSC-derived astrocyte/NVU perturbation experiments; transcriptomic cell-state signatures; imaging readouts; prospective validation & Causal support for interpreting genetic priors as vulnerability or response parameters & Stress--recovery kinetics, depletion slopes, persistent loss of recoverable capacity and observation maps & Large cohorts can calibrate useful priors but cannot by themselves prove that genetic susceptibility causally alters a specific depletion coefficient. Stronger interpretation requires triangulation with experimental, molecular, imaging and prospective evidence. \\ \addlinespace[2pt]
\end{longtable}
\end{footnotesize}
\end{landscape}

\clearpage
\section*{Supplementary Note 3 $|$ Mapping Alzheimer's disease hypotheses to PIM-BrainTwin model components}

This note expands the main-text description of PIM-BrainTwin as an open, modular framework in which alternative mechanistic accounts can be formalised, compared and iteratively refined. Alzheimer's disease hypotheses differ in their proposed initiating mechanisms, dominant biological scales and therapeutic targets. Contemporary mechanistic reviews describe amyloid, tau, neuroinflammation, oxidative stress, cholinergic dysfunction, glutamatergic excitotoxicity, metal-ion dyshomeostasis, microbiota--gut--brain signalling and impaired autophagy as major explanatory frameworks for Alzheimer's disease pathogenesis.\sfcite{1--3} PIM-BrainTwin does not treat these hypotheses as mutually exclusive narratives. Instead, each is translated into explicit state variables, interaction terms, parameter priors, observation equations and intervention handles within a common state-space architecture.

This translation has three advantages. First, it prevents hypothesis labels from substituting for model structure: an amyloid-centred, inflammatory-centred or vascular-centred account must specify which states, parameters and interactions are altered. Second, it allows competing model structures to be evaluated using shared criteria, including trajectory reconstruction, forecasting performance, sensitivity analysis, biological plausibility, perturbation response and model-credibility assessment. Third, it supports hybrid formulations in which several mechanisms contribute with different weights across individuals, disease stages and biological contexts.

The mapping below is intentionally conservative.\sfcite{4,10--16} It does not require every hypothesis to be represented as an independent state variable. Some mechanisms modify an existing state or parameter, some alter interactions between states, some affect observation maps, and others enter as exogenous forcing terms when suitable data are available. The glial-centred realisation developed in the main text is therefore one testable instantiation of the broader framework rather than a restriction on its mechanistic scope.

\begin{landscape}
\begin{footnotesize}
\setlength{\tabcolsep}{4pt}\renewcommand{\arraystretch}{1.25}
\begin{longtable}{>{\raggedright\arraybackslash}p{0.120\linewidth}>{\raggedright\arraybackslash}p{0.230\linewidth}>{\raggedright\arraybackslash}p{0.150\linewidth}>{\raggedright\arraybackslash}p{0.230\linewidth}>{\raggedright\arraybackslash}p{0.180\linewidth}}
\caption*{\textbf{Supplementary Table 3 $|$ Alzheimer's disease hypotheses and their representation within PIM-BrainTwin}}\\
\toprule
\textbf{Hypothesis or mechanism} & \textbf{Representation within PIM-BrainTwin} & \textbf{Primary state variables or layers} & \textbf{Candidate observations} & \textbf{Modelling role} \\
\midrule\endfirsthead
\multicolumn{5}{l}{\footnotesize\emph{Supplementary Table 3 (continued)}}\\[2pt]
\toprule
\textbf{Hypothesis or mechanism} & \textbf{Representation within PIM-BrainTwin} & \textbf{Primary state variables or layers} & \textbf{Candidate observations} & \textbf{Modelling role} \\
\midrule\endhead
\bottomrule\endfoot
Amyloid hypothesis & Amyloid production, aggregation, clearance and downstream interactions & $P_t$, with effects on $A_t$, $I_t$ and $V_t$ & Amyloid PET, CSF or plasma A$\beta$, APP/PSEN priors where relevant & Pathology forcing, clearance and pathology-to-capacity interactions \\ \addlinespace[2pt]
Tau hypothesis & Tau phosphorylation, aggregation, regional accumulation and network-mediated spread & $P_t$, $F_t$ and $M_t$ & Tau PET, p-tau, atrophy, FDG-PET, EEG or fMRI & Propagation, neuronal injury and downstream functional-network effects \\ \addlinespace[2pt]
Neuroinflammation & Glial activation, immune feedback, cytokine signalling and impaired inflammatory resolution & It, with effects on $A_t$ and $P_t$ & TSPO-PET, sTREM2, cytokines, GFAP, inflammatory proteomics and microglia-enriched transcriptomic signatures & Depletion pathway, feedback amplification and immune-resolution dynamics \\ \addlinespace[2pt]
Oxidative--metabolic stress & Mitochondrial dysfunction, impaired glucose metabolism, oxidative injury and reduced repair capacity & $M_t$ and the metabolic repair gate on $A_t$ & FDG-PET, metabolic markers, oxidative-stress markers and mitochondrial readouts & Repair limitation, metabolic vulnerability and fatigue-like depletion \\ \addlinespace[2pt]
Vascular--clearance dysfunction & BBB dysfunction, perfusion loss, impaired perivascular clearance, AQP4/endfoot disruption and haemodynamic loading & $V_t$, $A_t$ and clearance-related parameters & ASL-MRI, cerebrovascular reactivity, BBB markers, AQP4/endfoot measures and vascular-risk measures & Transport, clearance, compliance and mechanical or haemodynamic loading \\ \addlinespace[2pt]
Glutamatergic excitotoxicity & Synaptic stress, calcium-mediated toxicity and rapid network destabilisation & $F_t$ and $M_t$ & EEG, fMRI, synaptic markers, cognitive fluctuation and pharmacological response where available & Rapid perturbation of network physiology and metabolic stress \\ \addlinespace[2pt]
Cholinergic dysfunction & Modulation of attention and functional expression & Observation map from $F_t$ to $C_t$ & Cognitive testing, speech-derived markers, attentional measures and cholinergic-treatment response & Modifier of functional expression rather than a primary slow latent state \\ \addlinespace[2pt]
Metal-ion dyshomeostasis & Modulation of oxidative stress, protein aggregation and mitochondrial toxicity & $M_t$ and $P_t$ & Metal-sensitive biomarkers or imaging where available and oxidative-stress markers & Optional modifier of metabolic stress, oxidative injury and pathology aggregation \\ \addlinespace[2pt]
Autophagy and proteostasis & Impaired intracellular clearance, lysosomal dysfunction and proteostatic stress & $P_t$, $I_t$ and $M_t$ & Autophagy or lysosomal markers, transcriptomics and proteomics & Intracellular clearance, proteostasis and pathology-handling mechanisms \\ \addlinespace[2pt]
Microbiota--gut--brain axis & Peripheral inflammatory and metabolic forcing through immune, endocrine, microbial and metabolic pathways & Exogenous input to $I_t$ and $M_t$ & Microbiome profiles, systemic inflammatory markers, metabolomics and gut-barrier markers & Optional peripheral forcing term where data support model inclusion \\ \addlinespace[2pt]
Synaptic and neuronal injury & Synaptic dysfunction, neuronal loss and downstream network failure & $M_t$ and $F_t$, observed through $C_t$ & Structural MRI, FDG-PET, EEG or fMRI, synaptic markers, cognition and speech & Downstream injury pathway linking pathology and support-state failure to functional expression \\ \addlinespace[2pt]
Cognitive reserve & Preserved functional expression despite altered physiological state, potentially supported in part by neuroglial maintenance and adaptation & Observation layer $C_t$, reserve-related parameters and coupling between $A_t$ and $F_t$ & Education, occupational complexity, cognitive testing, speech and behavioural measures & Modifies the observation map from $F_t$ to cognition, speech and behaviour without introducing reserve as an additional primary latent state \\ \addlinespace[2pt]
\end{longtable}
\end{footnotesize}
\end{landscape}

\clearpage
\section*{Supplementary Note 4 $|$ Data capture, secure model updating and platform extensibility}

This note expands the main-text description of PIM-BrainTwin as an open, modular and auditable framework by specifying how the system captures data contributed by others, maintains accuracy, constructs and updates prediction models, and grows over time, while remaining scientifically rigorous and privacy-preserving.

\textbf{Data capture and integration.} We envision PIM-BrainTwin as a federated, version-controlled computational platform in which scientific extensibility, privacy and security are enforced at the architectural level. For each modality---including imaging, fluid biomarkers, transcriptomics, electrophysiology, speech and cognition---a publicly accessible, versioned contract specifies the data schema, metadata and observation function through which measurements are mapped onto one or more latent states, drawing where appropriate on the Brain Imaging Data Structure (BIDS) and GA4GH standards for interoperable data representation, federated access and machine-actionable data-use governance.\sfcite{36--38}

Raw clinical data never leave the boundaries of the contributing research centre. New data enter exclusively through explicitly defined observation models that map each modality onto one or more latent state variables. These local observation models are implemented as versioned, stateless computational functions. They are downloaded from the remote domain, executed only after explicit consent of the local centre, and run inside restricted, secure sandboxes. New observation models can be registered only through a predefined protocol that emphasises security review and validation.

To avoid repeated computation on resource-limited local machines, only privacy-protected latent-state representations (for example, through differential privacy), trained model weights and associated metadata---including version, uncertainty, consent records, access permissions and interoperability requirements---are uploaded to a centralised secure registry. These derived representations, model weights and metadata are the only data that leave the local site and are shared among authorised parties. Population-level aggregation, model fitting and computationally intensive simulation are performed exclusively on controlled PIM-BrainTwin high-performance computing infrastructure.

\textbf{Model credibility and quality control.} Model credibility is supported by the combination of physical and biological constraints, hierarchical Bayesian updating, and formal verification, validation and uncertainty-quantification requirements. Every observation model and every model release must pass numerical verification, uncertainty-calibration tests and locked temporal or external validation before it can affect population-level components of the framework. Prediction models are trained or updated from distributed data through the federated layer, with candidate model updates incorporated only after prespecified calibration and temporal or external validation.

\textbf{Growth of modelling capacity.} Growth proceeds along three explicitly separated dimensions.

\begin{itemize}[leftmargin=1.4em,itemsep=4pt]
\item \textbf{Dimension 1 -- Additional data within an existing centre.} New participant observations update individual posterior states, parameters and forecasts under a fixed model release. Individual updating remains completely separate from revision of the scientific framework. Population priors, observation functions, scale-transfer relationships and mechanistic modules change only through versioned releases that have passed the full verification and validation pipeline.
\item \textbf{Dimension 2 -- Accession of a new research centre.} A new centre first specifies the data types and observations it can provide, then runs the corresponding approved observation models inside secure local sandboxes and uploads only protected latent-state representations and metadata. It may optionally propose updated model weights derived from the newly contributed data; any such proposal is subject to the same verification and validation process before incorporation.
\item \textbf{Dimension 3 -- Introduction of a new observation model.} A new observation model is registered via the secure protocol. Participating centres must explicitly consent to its local execution. Once approved, the resulting derived representations flow into the central system under the same metadata and access-control regime.
\end{itemize}

\textbf{Feasibility.} The architecture builds on a federated-learning paradigm introduced by Google researchers in 2017\sfcite{39} and subsequently developed into mature frameworks and real-world multicentre biomedical applications, including Flower\sfcite{40,} federated-learning approaches in digital health,\sfcite{41} healthcare implementations led by Owkin\sfcite{42,} and cross-pharma federated learning at industrial scale.\sfcite{43,44} After nearly a decade of development, federated-learning technologies have progressed from methodological prototypes to real-world implementations in healthcare and pharmaceutical research. Industrial-scale modular cloud platforms for simulation and digital twins further demonstrate that distributed, version-controlled computational infrastructures are already operational.

From an engineering perspective, the principal challenges are the high flexibility required by scientific research and the stringent privacy and security constraints inherent to medical data. These challenges are substantial but manageable within a carefully designed federated, version-controlled architecture of the type described above.

This design allows PIM-BrainTwin to incorporate new data, research centres, observation models and predictive components while remaining privacy-preserving, auditable and scientifically constrained. It thereby enables the framework to evolve as new evidence, data sources and computational methods become available without conflating individual-level updating with revision of the shared scientific model.

\clearpage
\section*{Supplementary References}
\begin{refs}
\item De Strooper, B. \& Karran, E. The cellular phase of Alzheimer's disease. \emph{Cell} \textbf{164}, 603--615 (2016).
\item Knopman, D. S. et al. Alzheimer disease. \emph{Nat. Rev. Dis. Primers} \textbf{7}, 33 (2021).
\item Zhang, J. et al. Recent advances in Alzheimer's disease: mechanisms, clinical trials and new drug development strategies. \emph{Signal Transduct. Target. Ther.} \textbf{9}, 211 (2024).
\item Verkhratsky, A. \& Zorec, R. Neuroglia in cognitive reserve. \emph{Mol. Psychiatry} \textbf{29}, 3962--3967 (2024).
\item Green, G. S. et al. Cellular communities reveal trajectories of brain ageing and Alzheimer's disease. \emph{Nature} \textbf{633}, 634--645 (2024).
\item Gabitto, M. I. et al. Integrated multimodal cell atlas of Alzheimer's disease. \emph{Nat. Neurosci.} \textbf{27}, 2366--2383 (2024).
\item Miyoshi, E. et al. Spatial and single-nucleus transcriptomic analysis of genetic and sporadic forms of Alzheimer's disease. \emph{Nat. Genet.} \textbf{56}, 2704--2717 (2024).
\item Mathys, H. et al. Single-cell transcriptomic analysis of Alzheimer's disease. \emph{Nature} \textbf{570}, 332--337 (2019).
\item Grubman, A. et al. A single-cell atlas of entorhinal cortex from individuals with Alzheimer's disease reveals cell-type-specific gene expression regulation. \emph{Nat. Neurosci.} \textbf{22}, 2087--2097 (2019).
\item Abbott, N. J., Rönnbäck, L. \& Hansson, E. Astrocyte--endothelial interactions at the blood--brain barrier. \emph{Nat. Rev. Neurosci.} \textbf{7}, 41--53 (2006).
\item Sofroniew, M. V. \& Vinters, H. V. Astrocytes: biology and pathology. \emph{Acta Neuropathol.} \textbf{119}, 7--35 (2010).
\item Leng, F. \& Edison, P. Neuroinflammation and microglial activation in Alzheimer disease: where do we go from here? \emph{Nat. Rev. Neurol.} \textbf{17}, 157--172 (2021).
\item Zlokovic, B. V. The blood--brain barrier in health and chronic neurodegenerative disorders. \emph{Neuron} \textbf{57}, 178--201 (2008).
\item Butterfield, D. A. \& Halliwell, B. Oxidative stress, dysfunctional glucose metabolism and Alzheimer disease. \emph{Nat. Rev. Neurosci.} \textbf{20}, 148--160 (2019).
\item Korte, N., Nortley, R. \& Attwell, D. Cerebral blood flow decrease as an early pathological mechanism in Alzheimer's disease. \emph{Acta Neuropathol.} \textbf{140}, 793--810 (2020).
\item Iliff, J. J. et al. A paravascular pathway facilitates CSF flow through the brain parenchyma and the clearance of interstitial solutes, including amyloid $\beta$. \emph{Sci. Transl. Med.} \textbf{4}, 147ra111 (2012).
\item Backman, J. D. et al. Exome sequencing and analysis of 454,787 UK Biobank participants. \emph{Nature} \textbf{599}, 628--634 (2021).
\item Weiner, D. J. et al. Polygenic architecture of rare coding variation across 394,783 exomes. \emph{Nature} \textbf{614}, 492--499 (2023).
\item Satterstrom, F. K. et al. Large-scale exome sequencing study implicates both developmental and functional changes in the neurobiology of autism. \emph{Cell} \textbf{180}, 568--584.e23 (2020).
\item Li, M. et al. Integrative functional genomic analysis of human brain development and neuropsychiatric risks. \emph{Science} \textbf{362}, eaat7615 (2018).
\item Grotzinger, A. D. et al. Mapping the genetic landscape across 14 psychiatric disorders. \emph{Nature} \textbf{649}, 406--415 (2026).
\item Smeland, O. B. et al. A genome-wide analysis of the shared genetic risk architecture of complex neurological and psychiatric disorders. \emph{Nat. Neurosci.} \textbf{28}, 2439--2450 (2025).
\item Corder, E. H. et al. Gene dose of apolipoprotein E type 4 allele and the risk of Alzheimer's disease in late onset families. \emph{Science} \textbf{261}, 921--923 (1993).
\item Bellenguez, C. et al. New insights into the genetic etiology of Alzheimer's disease and related dementias. \emph{Nat. Genet.} \textbf{54}, 412--436 (2022).
\item Kunkle, B. W. et al. Genetic meta-analysis of diagnosed Alzheimer's disease identifies new risk loci and implicates A$\beta$, tau, immunity and lipid processing. \emph{Nat. Genet.} \textbf{51}, 414--430 (2019).
\item Jansen, I. E. et al. Genome-wide meta-analysis identifies new loci and functional pathways influencing Alzheimer's disease risk. \emph{Nat. Genet.} \textbf{51}, 404--413 (2019).
\item Yang, Z. et al. Cell-type-specific Alzheimer's disease polygenic risk scores are associated with distinct disease processes in Alzheimer's disease. \emph{Nat. Commun.} \textbf{14}, 7659 (2023).
\item Guerreiro, R. et al. TREM2 variants in Alzheimer's disease. \emph{N. Engl. J. Med.} \textbf{368}, 117--127 (2013).
\item Jonsson, T. et al. Variant of TREM2 associated with the risk of Alzheimer's disease. \emph{N. Engl. J. Med.} \textbf{368}, 107--116 (2013).
\item Sims, R. et al. Rare coding variants in PLCG2, ABI3, and TREM2 implicate microglial-mediated innate immunity in Alzheimer's disease. \emph{Nat. Genet.} \textbf{49}, 1373--1384 (2017).
\item Steinberg, S. et al. Loss-of-function variants in ABCA7 confer risk of Alzheimer's disease. \emph{Nat. Genet.} \textbf{47}, 445--447 (2015).
\item Raghavan, N. S. et al. Whole-exome sequencing in 20,197 persons for rare variants in Alzheimer's disease. \emph{Ann. Clin. Transl. Neurol.} \textbf{5}, 832--842 (2018).
\item Goate, A. et al. Segregation of a missense mutation in the amyloid precursor protein gene with familial Alzheimer's disease. \emph{Nature} \textbf{349}, 704--706 (1991).
\item Sherrington, R. et al. Cloning of a gene bearing missense mutations in early-onset familial Alzheimer's disease. \emph{Nature} \textbf{375}, 754--760 (1995).
\item Levy-Lahad, E. et al. Candidate gene for the chromosome 1 familial Alzheimer's disease locus. \emph{Science} \textbf{269}, 973--977 (1995).
\item Gorgolewski, K. J. et al. The brain imaging data structure, a format for organizing and describing outputs of neuroimaging experiments. \emph{Sci. Data} \textbf{3}, 160044 (2016).
\item Rehm, H. L. et al. GA4GH: International policies and standards for data sharing across genomic research and healthcare. \emph{Cell Genom.} \textbf{1}, 100029 (2021).
\item Lawson, J. et al. The Data Use Ontology to streamline responsible access to human biomedical datasets. \emph{Cell Genom.} \textbf{1}, 100028 (2021).
\item McMahan, H. B., Moore, E., Ramage, D., Hampson, S. \& Agüera y Arcas, B. Communication-efficient learning of deep networks from decentralized data. In \emph{Proc. 20th International Conference on Artificial Intelligence and Statistics} 1273--1282 (PMLR, 2017).
\item Beutel, D. J. et al. Flower: a friendly federated learning research framework. Preprint at \url{https://arxiv.org/abs/2007.14390} (2020).
\item Rieke, N. et al. The future of digital health with federated learning. \emph{npj Digit. Med.} \textbf{3}, 119 (2020).
\item Ogier du Terrail, J. et al. Federated learning for predicting histological response to neoadjuvant chemotherapy in triple-negative breast cancer. \emph{Nat. Med.} \textbf{29}, 135--146 (2023).
\item Oldenhof, M. et al. Industry-scale orchestrated federated learning for drug discovery. \emph{Proc. AAAI Conf. Artif. Intell.} \textbf{37}, 15576--15584 (2024).
\item Heyndrickx, W. et al. MELLODDY: cross-pharma federated learning at unprecedented scale unlocks benefits in QSAR without compromising proprietary information. \emph{J. Chem. Inf. Model.} \textbf{64}, 2331--2344 (2024).
\end{refs}